\documentclass[english,prb, twocolumn, showpacs]{revtex4}
\usepackage[T1]{fontenc}
\usepackage[latin9]{inputenc}
\usepackage{verbatim}
\usepackage{amsmath}
\usepackage{color}
\usepackage{graphicx}
\usepackage{amssymb}

\makeatletter

\providecommand{\tabularnewline}{\\}

\usepackage{hyperref}

\usepackage{babel}
\makeatother

\begin{document}

\title{Dissipative phase-fluctuations in superconducting wires capacitively
coupled to diffusive metals}

\author{Alejandro M. Lobos}

\affiliation{DPMC-MaNEP, University of Geneva, 24 Quai Ernest-Ansermet CH-1211
Geneva, Switzerland.}

\author{Thierry Giamarchi}

\affiliation{DPMC-MaNEP, University of Geneva, 24 Quai Ernest-Ansermet CH-1211
Geneva, Switzerland.}

\date{04/15/2010}

\begin{abstract}
We study the screening of the Coulomb interaction in a quasi one-dimensional
superconductor given by the presence of either a one- or a two-dimensional
non-interacting electron gas. To that end, we derive an effective
low-energy phase-only action, which amounts to treating the Coulomb
and superconducting correlations in the random-phase approximation.
We concentrate on the study of dissipation effects in the superconductor,
induced by the effect of Coulomb coupling to the diffusive density-modes
in the metal, and study its consequences on the static and dynamic
conductivity. Our results point towards the importance of the dimensionality
of the screening metal in the behavior of the superconducting plasma
mode of the wire at low energies. In absence of topological defects,
and when the screening is given by a one-dimensional electron gas,
the superconducting plasma mode is completely damped in the limit
$q\rightarrow0$, and consequently superconductivity is lost in the
wire. In contrast, we recover a Drude-response in the conductivity
when the screening is provided by a two-dimensional electron gas.
\end{abstract}


\maketitle

\section{Introduction}

%
{}

The environment has profound effects on the properties of quantum
systems \citep{caldeira_leggett}. %
{}In the case of superconductors, it was predicted more than 25 years
ago that a resistively shunted Josephson junction would experience
a superconductor-normal transition as a function of $R_{S}/R_{Q}$,
where $R_{S}$ is the shunt resistance of the junction and $R_{Q}=h/4e^{2}\approx6.45\;\text{k}\Omega$
is the quantum of resistance\citep{chakravarty82_macroscopic_quantum_tunneling,bray82_macroscopic_quantum_tunneling,schmid_instanton}.
%
{}More recently, a variety of superconducting systems, including granular\citep{Orr86_Global_phase_coherence_in_two-dimensional_granular_superconductors}
, or homogeneous\citep{Haviland89_Onset_of_superconductivity_in_the_two-dimensional_limit_PhysRevLett.62.2180}
films, 2D Josephson junctions arrays\citep{Geerligs89_Charging_effects_and_quantum_coherence_in_regular_Josephson_junction_arrays_PhysRevLett.63.326},
out-of-equilibrium Josephson junctions \citep{dallatorre09_quantum_criticality_non_equilibrium_noise}
and high temperature superconductors\citep{Sun94_Electron_tunneling_and_transport_in_HTSCs_PhysRevB.50.3266}
were shown to undergo a superconductor-insulator transition as the
characteristic resistance of the system in the normal state increases
through a critical value on the order of $R_{Q}$. In those cases,
the dissipative environment corresponds to the measurement circuits
or the intrinsic component of normal electrons in the system.

%
{}

In contrast, isolated superconducting wires with lateral dimension
$r_{0}\ll\xi_{0}$, where $\xi_{0}$ is the bulk coherence length,
do not present significant dissipation sources at low temperatures.
%
{}The low-energy modes in an ideally isolated superconducting wire are
the one-dimensional propagating plasmon modes along the axis \citep{mooij85_mooij_schon_mode}.
Contrary to bulk superconductors, where the plasmon has an energy
$\omega_{p}^{\textrm{3D}}=\sqrt{4\pi n_{s}e^{2}/m}$ (where $n_{s}$
is the superfluid density and $m$ is the electron mass), in the restricted
1D geometry of the wire, the long-range Coulomb interaction is not
completely screened and consequently charge fluctuations are not shifted
to finite energies in the limit $q\rightarrow0$. The result is a
sound-like dispersion relation $\omega^{2}\left(q\right)\sim q^{2}\ln\left(1/qr_{0}\right)$,
where the logarithmic factor is a remnant of the long-range Coulomb
interactions.

Because of the gapless dispersion relation, quantum fluctuations are
expected to show critical behavior\citep{giamarchi_book_1d}, a feature
that has attracted the attention of several theoretical \citep{zaikin97,buchler04_sit_finite_length_wire,goswami06_josephson_array,refael07_SN_transition_in_grains&nanowires}
and experimental\citep{bezryadin00,lau01,altomare06_experimental_evidence_of_qps}
research groups. %
{}

\begin{figure}[h]
\includegraphics[bb=50bp 360bp 550bp 690bp,clip,scale=0.5]{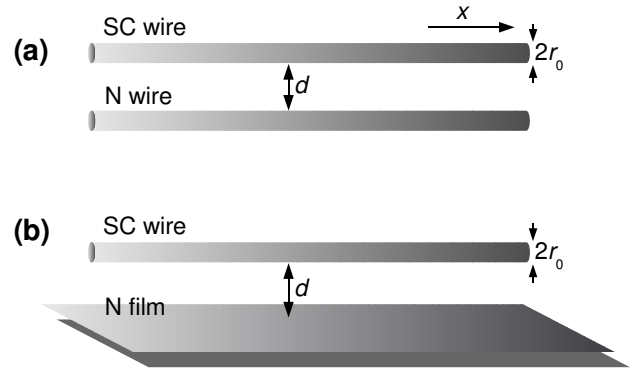}

\caption{\label{fig:system}Representation of the capacitively coupled superconducting
wire-normal metal system. The metal placed at a distance $d$ screens
the long-range Coulomb interaction in the superconducting wire. In
(a) the metal is a diffusive wire, and in (b) we consider a diffusive
2D electron gas .}

\end{figure}
How this picture (i.e., sound-like dispersion relation and critical
behavior) is modified when the coupling to the electromagnetic environment
is taken into account? Intuitively, the presence of a metal at a distance
$d$ should screen the long-range Coulomb interaction for density
fluctuations with wavelength $q\ll d^{-1}$, resulting in enhanced
superconducting correlations\citep{giamarchi_book_1d}%
{}. On the other hand, in capacitively-coupled superconductor-normal
systems, the presence of dissipation in the normal metal is known
to produce dissipative order-parameter fluctuations\citep{gaitonde98_dissipative_phase_fluctuations_2D,DePalo99_effective_action_BCS_BEC_crossover,rimberg97_dissipation_driven_sit_2D_josephson_array}
and, from this point of view, screening might also carry detrimental
effects to superconductivity. Moreover, recent theoretical works on
related Luttinger-liquid systems coupled electrostatically to metals
predict charge-density wave (CDW) instabilities caused by the dissipative
environment\citep{Gonzalez05_Coulomb_screening_and_electronic_instabilities_of_small-diameter_nanotubes_PhysRevB.72.205406,cazalilla06_dissipative_transition}.

Therefore, a better understanding of the screening effects occurring
in superconducting wires and the consequences to their superconducting
properties is needed. This issue is particularly relevant to recent
theoretical\citep{buchler04_sit_finite_length_wire,fu06_stabilization_of_superconductivity_in_nanowires_by_dissipation,lobos09_dissipation_scwires}
and experimental\citep{Liu09_Gate-Tunable_Dissipation_and_Superconductor-Insulator_Transition_in_Carbon_Nanotube_Josephson_Junctions,Chen09_Magnetic-Field-Induced_Superconducting_State_in_Zn_Nanowires_Driven_in_the_Normal_State_by_an_Electric_Current}
works showing evidence of stabilization of superconductivity in low
dimensional systems due to the presence of tunnelling contacts with
normal metallic leads, which suppress of fluctuations of the superconducting
order-parameter.

%
{}

%
{}%
{}

%
{}

%
{}

%
{}%
{}

In this article we study the effects of the screening of the Coulomb
interaction in a quasi-1D superconductor by the presence of a metal
nearby (cf. Fig. \ref{fig:system})%
{}. To that end, we derive a phase-only action of the coupled system
valid at low energies, which amounts to performing an RPA-approximation
of the interacting problem. We specify two experimentally relevant
geometries, namely: a) a 1D and b) a 2D electron gas (1DEG and 2DEG,
respectively) in the diffusive limit. Our results show a rich behavior
of the 1D plasma mode in the wire due to screening effects, and point
towards the importance of the dimensionality of the screening metal.
In particular in the case of screening provided by a 1DEG important
frictional effects are observed in the superconductor due to the capacitive
coupling, and in the limit $q\rightarrow0$ and $T\rightarrow0$ phase-coherence
is destroyed and the wire shows finite resistivity. In contrast, for
a wire screened by a 2DEG, friction and dissipation vanish in the
limit $q\rightarrow0$, and the wire is well described by the Luttinger
liquid picture.

The article is divided as follows: in Sec. \ref{sec:Model} we derive
a general effective phase-only action for the complete superconductor-normal
system, in Sec. \ref{sec:Screening_regimes} we present an analysis
of the screening regimes at low energies for both the 1D and 2D geometries,
Sec. \ref{sec:dynamic_conductivity} is devoted to the study of the
dissipative effects in the dynamical conductivity $\sigma\left(q,\omega\right)$
of the wire, and finally in Sec. \ref{sec:Discussion} we summarize
our findings and present a discussion. The details of the derivation
of the low-energy effective action are given in the Appendices \ref{sec:Derivation_S_eff}
and \ref{sec:Fourier_Transforms}.

\section{\label{sec:Model}Model}

In this section we derive a general effective model which describes
a clean superconductor capacitively coupled to a diffusive metal.
We leave for Sec. \ref{sec:Screening_regimes} the specific analysis
of the systems depicted in Fig. \ref{fig:system}, representing a
superconducting wire of length $L$ and lateral dimensions $r_{0}\ll\xi_{0}$
coupled to a diffusive metal placed at a distance $d$. The derivation
of the model is standard\citep{ambegaokar82_josephson_dissipation,zaikin97,vanotterlo98,DePalo99_effective_action_BCS_BEC_crossover}
and here we only sketch the main steps. We refer the reader to the
Appendix \ref{sec:Derivation_S_eff} and to the aforementioned references
for details.

In the following we use the convention $\hbar=k_{B}=1$. We begin
our description with the microscopic action of the complete system
\begin{align}
S & =\int_{0}^{\beta}d\tau\sum_{a,\sigma}\int d\mathbf{r}\;\psi_{a,\sigma}^{*}\left(\partial_{\tau}-\mu\right)\psi_{a,\sigma}+\int_{0}^{\beta}d\tau\; H,\label{eq:S_total}\end{align}
where $\beta=\frac{1}{T}$. The Grassmann field $\psi_{a,\sigma}\equiv\psi_{a,\sigma}\left(\mathbf{r},\tau\right)$
describes an electron in the superconductor for $a=s$ (normal metal
for $a=n$) with spin projection $\sigma$ at position $\mathbf{r}\equiv\left(x,y,z\right)$
and imaginary-time $\tau$. The chemical potential $\mu=k_{\text{F}}^{2}/2m$
is the Fermi energy in the normal state, with $k_{\text{F}}$ the
Fermi wavevector. The Hamiltonian $H$ of the systems is \begin{align}
H & =H_{\text{s}}^{0}+H_{\text{n}}^{0}+H_{\text{int}},\label{eq:h_total}\end{align}
where \begin{align}
H_{\text{s}}^{0} & =\int d\mathbf{r}\;\sum_{\sigma}\frac{\left[\nabla\psi_{s,\sigma}^{\dagger}\right]\left[\nabla\psi_{s,\sigma}\right]}{2m}+U\ \bar{\psi}_{s\uparrow}\bar{\psi}_{s\downarrow}\psi_{s\downarrow}\psi_{s\uparrow},\label{eq:h_s}\end{align}
describes a translationally invariant, clean superconductor. Since
we will not focus on the details of the pairing mechanism, here we
assume a phenomenological local attractive interaction $U<0$ which
is responsible for (s-wave) pairing at $T<T_{c}$.

The normal metal is described by \begin{align}
H_{\text{n}}^{0} & =\int d\mathbf{r}\;\sum_{\sigma}\left\{ \frac{\left[\nabla\psi_{n,\sigma}^{\dagger}\right]\left[\nabla\psi_{n,\sigma}\right]}{2m}+\psi_{n,\sigma}^{\dagger}V_{\text{i}}\psi_{n,\sigma}\right\} ,\label{eq:h_n}\end{align}
where $V_{\text{i}}\equiv V_{\text{i}}\left(\mathbf{r}\right)$ represents
the weak static impurity potential which provides a finite resitivity
in the metal.

Finally, the interaction term of the whole system is given by

\begin{align}
H_{\text{int}} & =\frac{1}{2}\int d\mathbf{r}_{1}d\mathbf{r}_{2}\;\hat{\rho}_{s}\left(\mathbf{r}_{1}\right)v\left(\mathbf{r}_{1}-\mathbf{r}_{2},0\right)\hat{\rho}_{s}\left(\mathbf{r}_{2}\right)+\nonumber \\
 & +\frac{1}{2}\int d\mathbf{r}_{1}d\mathbf{r}_{2}\;\hat{\rho}_{n}\left(\mathbf{r}_{1}\right)v\left(\mathbf{r}_{1}-\mathbf{r}_{2},0\right)\hat{\rho}_{n}\left(\mathbf{r}_{2}\right)+\nonumber \\
 & +\int d\mathbf{r}_{1}d\mathbf{r}_{2}\;\hat{\rho}_{s}\left(\mathbf{r}_{1}\right)v\left(\mathbf{r}_{1}-\mathbf{r}_{2},d\right)\hat{\rho}_{n}\left(\mathbf{r}_{2}\right),\label{eq:h_int_original}\end{align}
where we defined the electronic density operators $\hat{\rho}_{a}\left(\mathbf{r}\right)\equiv\sum_{\sigma}\psi_{a,\sigma}^{\dagger}\left(\mathbf{r}\right)\psi_{a,\sigma}\left(\mathbf{r}\right)$,
and where the domain of integration of the variables $\mathbf{r}_{1}$
and $\mathbf{r}_{2}$ is constrained to the volume of the superconductor
(if $a=s$) and the metal (if $a=n$). The interaction potential $v\left(\mathbf{r},z\right)$
is the microscopic long-range Coulomb interaction

\begin{align*}
v\left(\mathbf{r},z\right) & =\frac{1}{\epsilon_{\text{r}}}\frac{e^{2}}{\sqrt{r^{2}+z^{2}}},\end{align*}
where $\epsilon_{\text{r}}$ is the dielectric constant of the insulating
medium between the metal and the superconductor.

The first step in the derivation of an effective low-energy model
consists in decoupling the interaction terms appearing in $H_{\text{s}}^{0}$
and $H_{\text{int}}$ by the means of suitable Hubbard-Stratonovich
transformations (HSTs). The repulsive Coulomb interaction $H_{\text{int}}$
is more conveniently decoupled by expressing it in terms of the symmetric
and antisymmetric density operators\begin{align}
\hat{\rho}_{\pm}\left(\mathbf{r}\right) & \equiv\hat{\rho}_{s}\left(\mathbf{r}\right)\pm\hat{\rho}_{n}\left(\mathbf{r}\right).\label{eq:rho_nu}\end{align}
%
{}With this definition, the interaction term {[}cf. Eq. (\ref{eq:h_int_original})]
compactly writes \begin{align}
H_{\text{int}} & =\frac{1}{2}\sum_{\nu=\pm}\int d\mathbf{r}_{1}d\mathbf{r}_{2}\;\hat{\rho}_{\nu}\left(\mathbf{r}_{1}\right)v_{\nu}\left(\mathbf{r}_{1}-\mathbf{r}_{2}\right)\hat{\rho}_{\nu}\left(\mathbf{r}_{2}\right),\label{eq:h_int_nu}\end{align}
where we have defined\begin{align}
v_{\nu}\left(\mathbf{r}\right) & \equiv\frac{v\left(\mathbf{r},0\right)+\left(\nu\right)v\left(\mathbf{r},d\right)}{2}\quad\left(\text{with }\nu=\pm\right).\label{eq:v_nu}\end{align}
The HSTs to decouple the long-range Coulomb and the Hubbard $U<0$
interactions are implemented by introducing the HS fields $\tilde{\rho}_{\nu}\left(\mathbf{r},\tau\right)$
in the particle-hole channel, and $\bar{\Delta}\left(\mathbf{r},\tau\right),\Delta\left(\mathbf{r},\tau\right)$
in the particle-particle channel, respectively (cf. Appendix \ref{sec:Derivation_S_eff}).

The next step in our derivation is to introduce an extra HS field
$\rho_{\nu}\left(\mathbf{r},\tau\right)$ in order to decouple the
quadratic term in $\tilde{\rho}_{\nu}\left(\mathbf{r},\tau\right)$,
appearing in Eq. (\ref{eq:hs_ph}). Then, it is easy to show that
the field $\tilde{\rho}_{\nu}\left(\mathbf{r},\tau\right)$ can be
formally integrated out, yielding a functional-delta function\citep{negele}
$\delta\left[\hat{\rho}_{\nu}\left(\mathbf{r},\tau\right)-\rho_{\nu}\left(\mathbf{r},\tau\right)\right]$.
As noted by De Palo \textit{et al}. \citep{DePalo99_effective_action_BCS_BEC_crossover},
this fact allows to interpret the new HS fields $\rho_{\nu}\left(\mathbf{r},\tau\right)$
as the \textit{physical density} of the problem, expressed in our
case in terms of the symmetric and antisymmetric collective modes.

At sufficiently low energies, amplitude fluctuations of the order
parameter $\Delta\left(\mathbf{r},\tau\right)$ can be neglected,
allowing to write $\Delta\left(\mathbf{r},\tau\right)=\Delta_{0}e^{i\theta\left(\mathbf{r},\tau\right)}$,
with a real constant $\Delta_{0}$. The phase field $\theta\left(\mathbf{r},\tau\right)$
can be absorbed by a unitary transformation of the fermionic field\begin{align*}
\psi_{s,\sigma}\left(\mathbf{r},\tau\right) & \rightarrow\psi_{s,\sigma}^{\prime}\left(\mathbf{r},\tau\right)=\psi_{s,\sigma}\left(\mathbf{r},\tau\right)e^{i\theta\left(\mathbf{r},\tau\right)/2}.\end{align*}

The derivation of the effective model proceeds with the integration
of the fermionic fields $\psi_{a,\sigma}$, and by expanding the resulting
bosonic action around the saddle-point in terms of the derivatives
of $\theta\left(\mathbf{r},\tau\right)$ and the density fluctuations
$\delta\tilde{\rho}_{\nu}\left(\mathbf{r},\tau\right),\delta\rho_{\nu}\left(\mathbf{r},\tau\right)$
{[}cf. Eqs. (\ref{eq:rho_tilde_nu_fluctuation}) and (\ref{eq:rho_nu_fluctuation})].
This expansion amounts to performing an RPA-approximation of the interacting
problem\citep{DePalo99_effective_action_BCS_BEC_crossover,mahan2000}.

The last step is to integrate the auxiliary field $\delta\tilde{\rho}_{\nu}\left(\mathbf{r},\tau\right)$,
which in the original representation of the density in terms $\delta\rho_{s}\left(\mathbf{r},\tau\right),\delta\rho_{n}\left(\mathbf{r},\tau\right)$
yields \begin{align}
S_{\text{eff}} & =\int d\mathbf{r}d\tau\;\frac{i}{2}\partial_{\tau}\theta\left(\mathbf{r},\tau\right)\rho_{s}\left(\mathbf{r},\tau\right)+\nonumber \\
 & +\frac{1}{2}\int\prod_{i=1}^{2}d\mathbf{r}_{i}d\tau_{i}\;\left[\mathcal{D}\left(\mathbf{r}_{1}-\mathbf{r}_{2},\tau_{1}-\tau_{2}\right)\times\right.\nonumber \\
 & \times\nabla\theta\left(\mathbf{r}_{1},\tau_{1}\right)\nabla\theta\left(\mathbf{r}_{2},\tau_{2}\right)+\nonumber \\
 & \left.+\boldsymbol{\delta\rho}^{\dagger}\left(\mathbf{r}_{1},\tau_{1}\right)\boldsymbol{V}\left(\mathbf{r}_{1}-\mathbf{r}_{2},\tau_{1}-\tau_{2}\right)\boldsymbol{\delta\rho}\left(\mathbf{r}_{2},\tau_{2}\right)\right],\label{eq:S_eff}\end{align}
where $\mathcal{D}\left(\mathbf{r},\tau\right)$ is the phase stiffness
of the superconductor {[}cf. Eq. (\ref{eq:sc_stiffness})] and \begin{align*}
\boldsymbol{\delta\rho}\left(\mathbf{r},\tau\right) & \equiv\left(\begin{array}{c}
\delta\rho_{s}\left(\mathbf{r},\tau\right)\\
\delta\rho_{n}\left(\mathbf{r},\tau\right)\end{array}\right),\\
\boldsymbol{V}\left(\mathbf{r},\tau\right) & \equiv\left(\begin{array}{cc}
\left[\chi_{0,s}\left(\mathbf{r},\tau\right)\right]^{-1} & 0\\
0 & \left[\chi_{0,n}\left(\mathbf{r},\tau\right)\right]^{-1}\end{array}\right)+\\
 & +\delta\left(\tau\right)\left(\begin{array}{cc}
v\left(\mathbf{r},0\right) & v\left(\mathbf{r},d\right)\\
v\left(\mathbf{r},d\right) & v\left(\mathbf{r},0\right)\end{array}\right),\end{align*}
where we have used the notation $\left[\chi_{0,a}\left(\mathbf{r},\tau\right)\right]^{-1}\equiv\frac{1}{\beta V}\sum_{\mathbf{k},\omega_{m}}e^{i\mathbf{k}.\mathbf{r}-i\omega_{m}\tau}\chi_{0,a}^{-1}\left(\mathbf{k},\omega_{m}\right)$,
where $\chi_{0,a}\left(\mathbf{k},\omega_{m}\right)$ is the bare
density-density correlator (i.e., obtained with the Hamiltonian $H_{a}^{0}$),
defined in Eqs. (\ref{eq:sc_susceptibility}) and (\ref{eq:n_susceptibility}).
Here we have used the notation in Fourier space $\left(\mathbf{k},\omega_{m}\right)$
with $\mathbf{k}$ the momentum and $\omega_{m}=\frac{2\pi m}{\beta}$
the bosonic Matsubara frequencies\citep{mahan2000}.%
{}

Note that at $T=0$ and in absence of quasiparticle excitations, the
whole electronic density in the superconductor corresponds to the
superfluid density. Consequently, the field $\delta\rho_{s}\left(\mathbf{r},\tau\right)$
physically represents the fluctuation of the Cooper-pair density at
point $\left(\mathbf{r},\tau\right)$.

An interesting aspect of the effective action in Eq. (\ref{eq:S_eff})
is that the first term (i.e., coupling between the total density of
Cooper-pairs $\rho_{s}\left(\mathbf{r},\tau\right)$ and the phase-field
$\theta\left(\mathbf{r},\tau\right)$) appears naturally as a consequence
of the well-known number-phase commutation-relation $\left[\rho_{s}\left(\mathbf{r}\right),\theta\left(\mathbf{r}^{\prime}\right)\right]=i\delta\left(\mathbf{r}-\mathbf{r}^{\prime}\right)$
occurring in the superconducting groundstate \citep{Tinkham}.

Besides the contribution of soft modes, encoded in Eq. (\ref{eq:S_eff}),
in low-dimensional superconductors there are also stable topological
excitations which contribute to the effective action. These are the
well-known classical vortex (in 2D) and the phase slips (in 1D) excitations\citep{Tinkham,langer67,mccumber70}.
Focusing in the 1D case, a phase-slip is a region of size $\sim\xi_{0}$
where the order parameter temporarily vanishes, allowing the field
$\theta\left(\mathbf{r},\tau\right)$ to perform a jump of $\pm2\pi n$
(with $n$ integer) across it. For wires in the limit of very low
superconducting stiffness, phase slips are an important source of
momentum-unbinding, and a relevant contribution to the action in the
RG-sense\citep{giamarchi_attract_1d,zaikin97,giamarchi_book_1d}.
Indeed, it is believed that the eventual destruction of the superconducting
state in isolated ultrathin wires occurs through the proliferation
of quantum phase slips/anti phase slips pairs\citep{giamarchi_attract_1d,giordano94,zaikin97,bezryadin00,lau01,buchler04_sit_finite_length_wire,altomare06_experimental_evidence_of_qps,arutyunov08_superconductivity_1d_review},
in what constitutes the quantum analog in 1+1 dimensions to the classical
Berezinskii-Kosterlitz-Thouless (BKT) transition in two space dimensions\citep{kosterlitz_thouless}.

Note that our derivation does not account for the presence of phase-slips.
Consequently, our results will only apply far from the BKT transition
and far from the (non-superconducting) phase where the effect of phase
slips dominates.

In the following we analyze the generic action of Eq.(\ref{eq:S_eff})
for the different configurations of Fig. \ref{fig:system}.

\section{\label{sec:Screening_regimes}Screening regimes}

\subsection{\label{sub:unscreened}Unscreened isolated wire}

Let us first explore the instructive case of a superconducting wire
ideally isolated from the environment. This situation corresponds
to the normal metal placed infinitely far from the superconductor
(i.e., $d\rightarrow\infty$), which results in the decoupling of
their dynamics. For a very narrow superconducting wire with $r_{0}\ll\xi_{0}$,
the dependence of the fields $\theta\left(\mathbf{r},\tau\right),\delta\rho_{s}\left(\mathbf{r},\tau\right)$
on transverse dimensions can be neglected, reducing to $\left\{ \theta\left(\mathbf{r},\tau\right),\delta\rho_{s}\left(\mathbf{r},\tau\right)\right\} \rightarrow\left\{ \theta\left(\mathbf{x}\right),\delta\rho_{s}\left(\mathbf{x}\right)\right\} $
where the compact notation $\mathbf{x}=\equiv\left(x,\tau\right)$
has been used. The effective action of the superconducting wire writes
most conveniently in Fourier space

\begin{align}
S_{0}^{\textrm{w}} & \simeq\frac{1}{2\beta L}\sum_{\mathbf{q}}\omega_{m}\theta\left(\mathbf{q}\right)\rho_{s}^{*}\left(\mathbf{q}\right)+q^{2}\mathcal{D}\left(\mathbf{q}\right)\left|\theta\left(\mathbf{q}\right)\right|^{2}+,\nonumber \\
 & +\frac{1+\chi_{0,s}\left(\mathbf{q}\right)v\left(q,0\right)}{\chi_{0,s}\left(\mathbf{q}\right)}\left|\rho_{s}\left(\mathbf{q}\right)\right|^{2},\label{eq:S_isolated}\end{align}
where we have used the notation $\mathbf{q}\equiv\left(q,-\omega_{m}\right)$
(with $q$ the momentum along the wire), and the property of real
fields $\theta^{*}\left(\mathbf{q}\right)=\theta\left(-\mathbf{q}\right)$,
$\rho_{s}^{*}\left(\mathbf{q}\right)=\rho_{s}\left(-\mathbf{q}\right)$.
The Fourier transforms $\mathcal{D}\left(\mathbf{q}\right)$ and $\chi_{0,s}\left(\mathbf{q}\right)$
are defined as

\begin{align}
\mathcal{D}\left(\mathbf{q}\right) & =\int_{0}^{\beta}d\tau\int_{0}^{L}dx\int dydz\; e^{-i\mathbf{q}\mathbf{x}}\;\mathcal{D}\left(\mathbf{r},\tau\right),\label{eq:FT_sc_stiffness}\\
\chi_{0,s}\left(\mathbf{q}\right) & =\int_{0}^{\beta}d\tau\int_{0}^{L}dx\int dydz\; e^{-i\mathbf{q}\mathbf{x}}\;\chi_{0,s}\left(\mathbf{r},\tau\right).\label{eq:FT_sc_susceptibility}\end{align}
At this point is relevant to calculate the Fourier transform of the
Coulomb potential, which can be approximated as $v\left(\mathbf{r},z\right)\simeq\frac{e^{2}}{\epsilon_{\text{r}}\sqrt{x^{2}+r_{0}^{2}+z^{2}}}$
(i.e., only dependent on the spatial coordinate $x$), and cut off
at short distances by the radius $r_{0}$. Therefore we have\begin{alignat}{1}
v\left(q,d\right) & =\frac{2e^{2}}{\epsilon_{\text{r}}}K_{0}\left(\left|q\right|\sqrt{r_{0}^{2}+d^{2}}\right),\label{eq:v_q_z}\end{alignat}
where $K_{0}\left(\zeta\right)$ is the zeroth-order Bessel function,
which verifies the limit $\lim_{\zeta\rightarrow0}K_{0}\left(\zeta\right)\rightarrow-\ln\left(\frac{\zeta}{2}\right)-\gamma,$
with $\gamma$ the Euler gamma constant\citep{abramowitz_math_functions}.
From the above Eq. (\ref{eq:S_isolated}) we obtain the phase-only
action in the limit $\mathbf{q}\rightarrow0$, by integration of the
field $\rho_{s}\left(\mathbf{q}\right)$ \begin{align}
S_{0}^{\text{w}} & \simeq\frac{1}{2\beta L}\sum_{\mathbf{q}}\left[\frac{\omega_{m}^{2}}{\frac{8e^{2}}{\epsilon_{\text{r}}}\ln\left(\frac{2}{\left|qr_{0}\right|}\right)}+q^{2}\mathcal{D}_{0}\right]\left|\theta\left(\mathbf{q}\right)\right|^{2},\label{eq:S_isolated_phase_only}\end{align}
where $\mathcal{D}_{0}\equiv\lim_{\mathbf{q}\rightarrow0}\:\mathcal{D}\left(\mathbf{q}\right)=\frac{\rho_{s}^{\left(0\right)}}{4m}$
(cf. Appendix \ref{sec:Fourier_Transforms}).

The minimization of the effective action Eq. (\ref{eq:S_isolated_phase_only})
allows to obtain the equation of motion for the phase-field and to
recover the dispersion-relation predicted for the 1D-plasma mode \citep{mooij85_mooij_schon_mode,giamarchi_book_1d}
upon analytical continuation to real frequencies $i\omega_{m}\rightarrow\omega+i0^{+}$\begin{align}
\omega^{2}\left(q\right)-\frac{8e^{2}}{\epsilon_{\text{r}}}\mathcal{D}_{0}q^{2}\ln\left(\frac{2}{\left|qr_{0}\right|}\right) & =0.\label{eq:dispersion_relation_mooij_schoen}\end{align}

%
{}Let us now concentrate on the superconducting properties of the wire.
It is well-known that long-range order of the order parameter in 1D
quantum systems is not possible, due to presence of strong quantum
fluctuations and, strictly speaking, only quasi-long-range order,
characterized by a slowly decreasing order-parameter correlation function\begin{align}
F\left(\mathbf{x}\right) & \equiv\left\langle \Delta^{*}\left(\mathbf{x}\right)\Delta\left(0\right)\right\rangle \nonumber \\
 & =\Delta_{0}^{2}e^{-\frac{1}{2}\left\langle T_{\tau}\left[\theta\left(\mathbf{x}\right)-\theta\left(0\right)\right]^{2}\right\rangle },\label{eq:sc_correlation_function}\end{align}
can exist \citep{mermin_wagner_theorem,giamarchi_book_1d}. In the
case of the isolated wire, the phase-correlation function calculated
with the effective phase-only action Eq. (\ref{eq:S_isolated_phase_only})
writes\citep{schulz_wigner_1d,giamarchi_book_1d}\begin{align}
\left\langle T_{\tau}\left[\theta\left(\mathbf{x}\right)-\theta\left(0\right)\right]^{2}\right\rangle  & =\frac{1}{\pi K}\left[\ln\left(\frac{\sqrt{x^{2}+u^{2}\tau^{2}\ln\tau}}{r_{0}}\right)\right]^{3/2},\label{eq:phase_correlator_wigner}\end{align}
{}where $K\equiv\sqrt{\frac{\mathcal{D}_{0}\epsilon_{\text{r}}}{8e^{2}}}$
and $u\equiv\sqrt{\frac{\mathcal{D}_{0}8e^{2}}{\epsilon_{\text{r}}}}$.
As compared with the case of a 1D superconductor with short-range
repulsive interactions\citep{giamarchi_book_1d}, the phase correlator
of Eq. (\ref{eq:phase_correlator_wigner}) produces a relatively fast
decrease of the order-parameter correlation function Eq. (\ref{eq:sc_correlation_function}),
as a consequence of the long-range Coulomb interaction, which is not
completely screened in the 1D geometry. Consequently, density fluctuations
are suppressed in the limit $\mathbf{q}\rightarrow0$\citep{schulz_wigner_1d},
and superconductivity, which benefits from fluctuations in the density,
is suppressed.

A natural step to take in order to diminish the detrimental effects
of the Coulomb interaction in the 1D geometry, is to screen it by
the means of a metal placed nearby. This is the subject of the subsequent
sections.

\subsection{\label{sub:screening1D}Screening by a diffusive metallic wire}

We now concentrate on the system depicted in Fig. \ref{fig:system}(a).
For simplicity, we consider the case of two geometrically identical
cylindrical wires. Extensions to other 1D geometries are straightforward.
We assume that the normal metal is only one-dimensional with respect
to density fluctuations $\rho_{n}\left(\mathbf{q}\right)$ with spatial
wavevector $q$ satisfying the condition $qr_{0}\ll1$. Note that
this condition does not necessarily imply that the normal wire is
\textit{electronically} 1D (i.e., it does not imply the existence
of only one electronic conduction channel). Indeed, in what follows
we assume a normal metal with a large number of channels $N_{\text{ch}}\sim\left(k_{\text{F}}r_{0}\right)^{2}\gg1$.
This fact, together with the additional assumption of a very weak
disorder potential, allows to neglect Anderson-localization effects
(i.e., $L\ll\xi_{\textrm{wire}}$, where $\xi_{\textrm{wire}}$ is
the localization length in the diffusive normal wire).

\begin{figure}[h]
\includegraphics[bb=0bp 0bp 390bp 230bp,clip,scale=0.65]{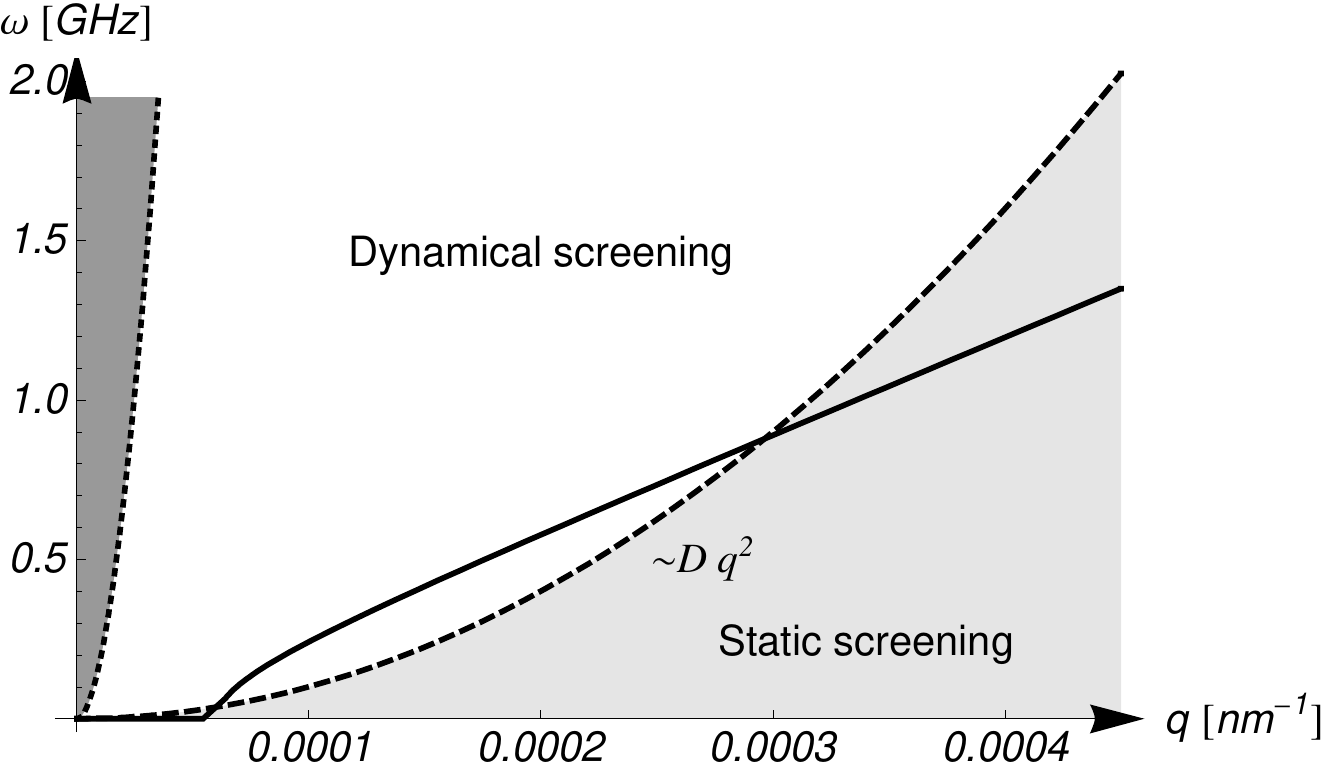}

\caption{\label{fig:screening_regime_1D}Screening regimes for a superconducting
wire screened by a diffusive 1DEG. The curve $\omega=Dq^{2}$ separates
the regime of static screening $\omega\ll Dq^{2}$ (light gray area)
from that of dynamical screening $\omega\gg Dq^{2}$ (white area).
The dispersion relation of the 1D plasma mode (thick solid line) is
obtained from the solution of Eq. (\ref{eq:S_eff_1D_phase_only}).
For typical experimental values (cf. Table \ref{tab:parameters}),
the dispersion relation crosses over from the static regime to the
dynamical regime, and eventually the mode is completely damped. The
unscreened regime of frequencies $\omega\gg Dq^{2}\frac{2e^{2}}{\epsilon_{\text{r}}}\mathcal{N}_{n,1D}^{0}\ln\frac{2}{qr_{0}}$
corresponds to the dark gray area.}

\end{figure}
In that case, Eq. (\ref{eq:S_eff}) reduces to \begin{align}
S_{\left(1\right)}^{\text{w}} & \simeq\frac{1}{2\beta L}\sum_{\mathbf{q}}\left[\omega_{m}\theta\left(\mathbf{q}\right)\rho_{s}^{*}\left(\mathbf{q}\right)+q^{2}\mathcal{D}\left(\mathbf{q}\right)\left|\theta\left(\mathbf{q}\right)\right|^{2}+\right.\nonumber \\
 & \left.+\boldsymbol{\rho}^{\dagger}\left(\mathbf{q}\right)\boldsymbol{V}\left(\mathbf{q}\right)\boldsymbol{\rho}\left(-\mathbf{q}\right)\right],\label{eq:S_eff_1D}\end{align}
where the subindex $g$ in $S_{\left(g\right)}^{\text{w}}$ indicates
the effective dimensionality of the metal. %
{}The integration of the density modes $\rho_{s}\left(\mathbf{q}\right)$
and $\rho_{n}\left(\mathbf{q}\right)$ in the above expression allows
to obtain the result%
{} \begin{align}
S_{\left(1\right)}^{\text{w}} & \simeq\frac{1}{2\beta L}\sum_{\mathbf{q}}\left\{ \frac{\omega_{m}^{2}}{4}\left[\frac{1+\Bigl(\chi_{0,s}\left(\mathbf{q}\right)+\chi_{0,n}\left(\mathbf{q}\right)\Bigr)v\left(q,0\right)}{\chi_{0,s}\left(\mathbf{q}\right)\left(1+\chi_{0,n}\left(\mathbf{q}\right)v\left(q,0\right)\right)}+\right.\right.\nonumber \\
 & \left.+\frac{\chi_{0,s}\left(\mathbf{q}\right)\chi_{0,n}\left(\mathbf{q}\right)\left(v\left(q,0\right)^{2}-v\left(q,d\right)^{2}\right)}{\chi_{0,s}\left(\mathbf{q}\right)\left(1+\chi_{0,n}\left(\mathbf{q}\right)v\left(q,0\right)\right)}\right]^{-1}+\nonumber \\
 & +q^{2}\mathcal{D}\left(\mathbf{q}\right)\biggr\}\left|\theta\left(\mathbf{q}\right)\right|^{2}.\label{eq:S_eff_1D_sc}\end{align}
%
{}

In the following we focus on the experimentally relevant regime $d\approx r_{0}\ll q^{-1}$.
In that case the quantity $v\left(q,0\right)^{2}-v\left(q,d\right)^{2}\sim\left[\ln\frac{r_{0}}{d}\right]^{2}\simeq0$
drops from Eq. (\ref{eq:S_eff_1D_sc}) and the expression simplifies
to%
{} \begin{align}
S_{\left(1\right)}^{\text{w}} & \simeq\frac{1}{2\beta L}\sum_{\mathbf{q}}\left\{ \frac{\omega_{m}^{2}}{4}\frac{\chi_{0,s}\left(\mathbf{q}\right)\left[1+\chi_{0,n}\left(\mathbf{q}\right)v\left(q,0\right)\right]}{1+\left[\chi_{0,s}\left(\mathbf{q}\right)+\chi_{0,n}\left(\mathbf{q}\right)\right]v\left(q,0\right)}+\right.\nonumber \\
 & \left.+q^{2}\mathcal{D}\left(\mathbf{q}\right)\right\} \left|\theta\left(\mathbf{q}\right)\right|^{2}.\label{eq:S_eff_1D_phase_only}\end{align}

\begin{figure}[h]
\includegraphics[bb=0bp 0bp 380bp 220bp,clip,scale=0.65]{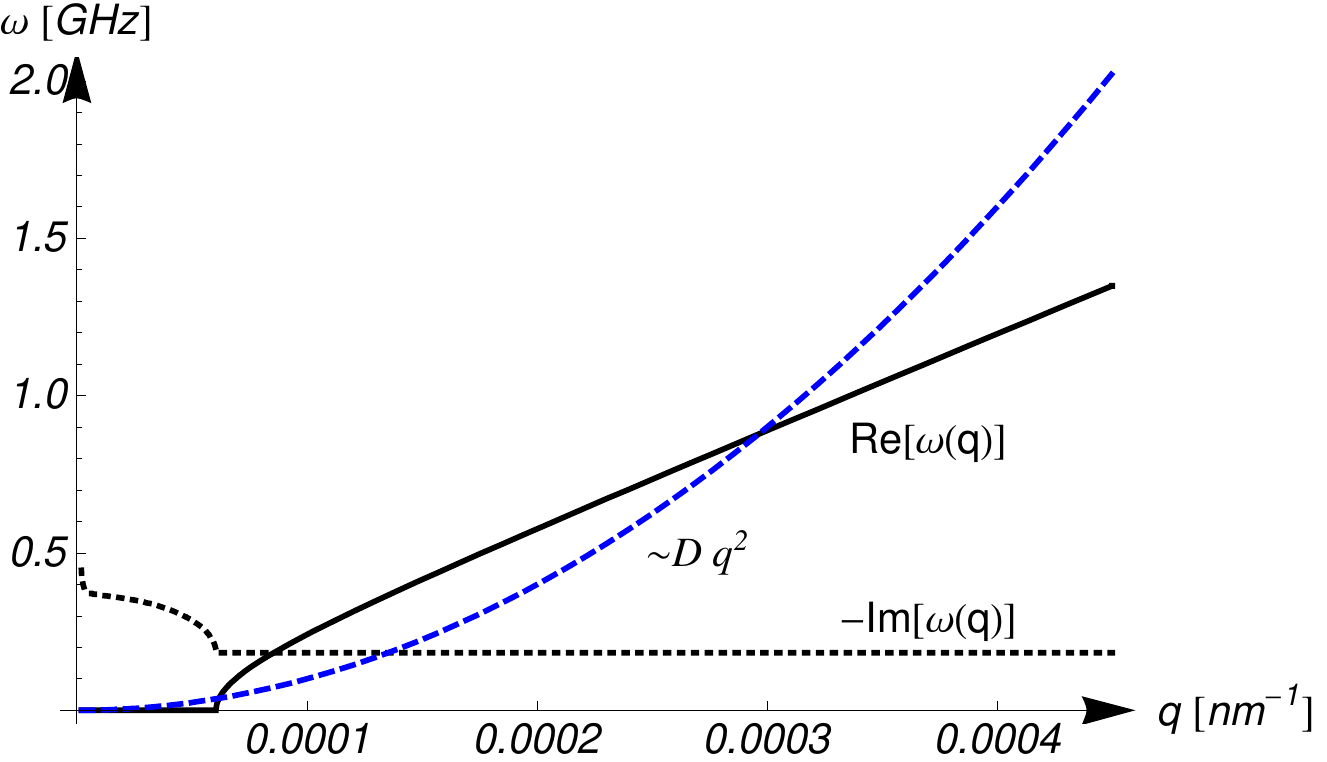}

\caption{\label{fig:im_real_1D}Real and imaginary components of the 1D plasma
mode $\omega\left(q\right)$, obtained from the equation of motion
of the action Eq. (\ref{eq:S_eff_1D_phase_only}). The real part (solid
line) gives the dispersion relation, while the imaginary part (dotted
line) represents the damping of the mode. As in Fig. \ref{fig:screening_regime_1D},
the curves have been calculated for realistic experimental parameters
(cf. Table \ref{tab:parameters}). The curve $Dq^{2}$ (blue dashed
line) is shown as a reference. }

\end{figure}
For a weakly-disordered diffusive electron gas with elastic mean-free
path $l_{e}$ and scattering time $\tau_{e}=l_{e}/v_{F}$, where $v_{F}$
is the Fermi velocity, the disorder-averaged density-density correlation
function {[}cf. Eq. (\ref{eq:n_susceptibility})] at energies $\left|\omega_{m}\right|<\tau_{e}^{-1}$
and momentum $q<l_{e}^{-1}$ writes \citep{akkermans}\begin{align}
\chi_{0,n}\left(\mathbf{q}\right) & \simeq2\mathcal{N}_{n,1D}^{0}\frac{Dq^{2}}{Dq^{2}+\left|\omega_{m}\right|},\label{eq:xi0n}\end{align}
where $\mathcal{N}_{n,1D}^{0}$ is the 1D density of states at the
Fermi energy in the normal metal, and $D=l_{e}^{2}/\tau_{e}$ is the
diffusion constant in 1D. The factor 2 accounts for the spin degeneracy.
%
{}

Note that the susceptibility $\chi_{0,n}\left(\mathbf{q}\right)$
{[}cf. Eq. (\ref{eq:xi0n})] is non-analytical in the limit $\mathbf{q}\rightarrow0$
for a normal diffusive metal. On the contrary, for the superconductor
the presence of a gap in the excitation spectrum allows to obtain
a well-defined limit $\lim_{\mathbf{q}\rightarrow0}\chi_{0,s}\left(\mathbf{q}\right)\simeq\gamma\mathcal{N}_{s,1D}^{0}$,
where $\mathcal{N}_{s,1D}^{0}$ is the linear density of states in
the superconductor (in the normal state) at the Fermi level, and $\gamma$
is a numerical coefficient of order 1 {[}cf. Eq. (\ref{eq:susceptibility_sc_limit_q_0})].

The plasma mode obtained from the equations of motion derived from
Eq. (\ref{eq:S_eff_1D_phase_only}) is plotted in Fig. \ref{fig:screening_regime_1D}
(thick solid line). Due to the complexity of the screening provided
by the diffusive 1DEG, it is instructive to derive analytical expressions
valid in the limiting cases of static (i.e., $\left|\omega_{m}\right|\ll Dq^{2}$)
and dynamical (i.e., $Dq^{2}\ll\left|\omega_{m}\right|$) screening.

%
{}

\subsubsection{Static screening limit $\left|\omega_{m}\right|\ll Dq^{2}$}

{}

This limit corresponds to the region $\left|\omega_{m}\right|\ll Dq^{2}$
(see light gray area in Fig. \ref{fig:screening_regime_1D}). In this
case, the susceptibility in the normal metal can be approximated as
$\chi_{0,n}\left(\mathbf{q}\right)\simeq2\mathcal{N}_{n}^{0}\left(1-\frac{\left|\omega_{m}\right|}{Dq^{2}}\right)$
{[}cf. Eq. (\ref{eq:xi0n})] . Then Eq. (\ref{eq:S_eff_1D_phase_only})
can be written as\begin{align}
S_{\left(1\right)}^{\textrm{w}} & \simeq\frac{1}{2\beta L}\sum_{\mathbf{q}}\left[\chi_{s}\left(0\right)\omega_{m}^{2}\left(1-\frac{\alpha\left|\omega_{m}\right|}{Dq^{2}}\right)+q^{2}\mathcal{D}_{0}\right]\left|\theta\left(\mathbf{q}\right)\right|^{2}\label{eq:S_eff_1D_static}\end{align}
with $\chi_{s}\left(0\right)\equiv\frac{2\gamma\mathcal{N}_{n,1D}^{0}\mathcal{N}_{s,1D}^{0}}{\gamma\mathcal{N}_{s,1D}^{0}+2\mathcal{N}_{n,1D}^{0}}$
the effective static RPA-susceptibility of the wire and $\alpha\equiv\frac{\gamma\mathcal{N}_{s,1D}^{0}}{\gamma\mathcal{N}_{s,1D}^{0}+2\mathcal{N}_{n,1D}^{0}}$.
Note that in the limit $\alpha\rightarrow0$, the above action corresponds
to a Luttinger liquid action with short-range interactions\citep{giamarchi_book_1d}.
In the case of a 1D geometry of Fig. \ref{fig:system}(a), the screening
length is given by the distance $d$.

In the more general case of $\alpha>0$, the term $\sim\frac{\alpha\left|\omega_{m}\right|}{Dq^{2}}$
introduces dissipation in the plasmon mode. %
{}%
{}%
{} From Eq. (\ref{eq:S_eff_1D_static}), the dispersion relation for
the plasma-mode writes \begin{align}
-\omega^{2}\left(q\right)\left(1+i\frac{\alpha\omega\left(q\right)}{Dq^{2}}\right)-\frac{\mathcal{D}_{0}}{\chi_{s}\left(0\right)}q^{2} & =0.\label{eq:LL_dispersion_relation}\end{align}
This equation holds provided the consistency condition $\left|\omega\left(q\right)\right|\ll Dq^{2}$
is verified (cf. solid line in Fig. \ref{fig:screening_regime_1D}).
In Fig. \ref{fig:im_real_1D} we show the solution of the above Eq.
(\ref{eq:LL_dispersion_relation}) as a function of $q$. Note that
while Re{[}$\omega\left(q\right)$] follows an approximately linear
dispersion relation, the imaginary part takes a constant value in
the regime $\left|\omega\left(q\right)\right|\ll Dq^{2}$, meaning
that the plasmon mode acquires a finite width, which in the perturbative
limit $\alpha\rightarrow0$ writes $\Gamma\left(q\right)\equiv-\text{Im}\left[\omega\left(q\right)\right]\simeq\frac{\alpha\mathcal{D}_{0}}{2D\chi_{s}\left(0\right)}$
(cf. Fig. \ref{fig:im_real_1D}).%
{}

\subsubsection{\label{sub:1D_screening_dynamic}Dynamic screening limit  $\left|\omega_{m}\right|\gg Dq^{2}$}

For realisitic estimates of the experimental parameters (cf. Table
\ref{tab:parameters}), our results indicate that the regime $\left|\omega_{m}\right|\gg Dq^{2}$
(white area in Fig. \ref{fig:screening_regime_1D}) is the most relevant
for experimental studies on today's accesible wires \citep{bezryadin00,lau01,altomare06_experimental_evidence_of_qps}.
Replacing Eq. (\ref{eq:xi0n}) into Eq. (\ref{eq:S_eff_1D_phase_only})
we note that  if the condition

\begin{align}
Dq^{2}\ll\left|\omega_{m}\right| & \ll\frac{2e^{2}}{\epsilon_{\textrm{r}}}\mathcal{N}_{n,1D}^{0}Dq^{2}\ln\frac{2}{qr_{0}},\label{eq:condition_dynamical_screening}\end{align}
is fulfilled, the action in Eq. (\ref{eq:S_eff_1D_phase_only}) can
be approximated as%
{}\begin{align}
S_{\left(1\right)}^{\textrm{w}} & \simeq\frac{1}{2\beta L}\sum_{\mathbf{q}}\left[2\mathcal{N}_{n,1D}^{0}Dq^{2}\left|\omega_{m}\right|+\mathcal{D}_{0}q^{2}\right]\left|\theta\left(\mathbf{q}\right)\right|^{2}.\label{eq:S_diff_theta}\end{align}
The action Eq. (\ref{eq:S_diff_theta}) indicates that phase fluctuations
show dissipative dynamics (encoded in the term $\sim q^{2}\left|\omega_{m}\right|$)
as a consequence of the coupling to the dissipative processes in the
1DEG. In other words, the superconductor {}``inherits'' the dissipation
in the 1DEG through the Coulomb interaction.

Note that a term $\sim q^{2}\left|\omega_{m}\right|$ has been studied
in the context of resistively shunted Josephson junctions arrays (RSJJAs)\citep{Chakavarty86_Onset_of_Global_Phase_Coherence_in_Josephson-Junction_Arrays,chakravarty88_rsjj_phase_diagram,goswami06_josephson_array}.
In that case, the term $\sim q^{2}\left|\omega_{m}\right|$ appears\textit{
in addition} to the dynamical term $\sim\omega_{m}^{2}$, which represents
the effect of quantum fluctuations induced by the charging energy
of the superconducting island\citep{fazio01_review_superconducting_networks}.
As a result, dissipation turns out to be beneficial to superconductivity,
through the quenching of phase fluctuations\citep{Chakavarty86_Onset_of_Global_Phase_Coherence_in_Josephson-Junction_Arrays}.

However, in our case, the form of the action in Eq. (\ref{eq:S_diff_theta})
is qualitatively different, since the term $\sim\omega_{m}^{2}$ is
\textit{absent} from the action (actually, it is the dynamical term
\textit{itself} which becomes a contribution $\sim q^{2}\left|\omega_{m}\right|$).
This has \textit{detrimental} consequences for the superconductivity
in the wire, as can be seen directly from the equation of motion for
the field $\theta$, which gives $\omega\left(q\right)\simeq-i\mathcal{D}_{0}/\left(D\mathcal{N}_{n,1D}^{0}\right)$,
indicating that the original plasma mode is completely damped and
vanishes in the limit $q\rightarrow0$ (see Fig. \ref{fig:screening_regime_1D}).
Indeed, expressing the action Eq. (\ref{eq:S_eff_1D}) in terms of
the dual field\citep{giamarchi_book_1d} $\phi\left(\mathbf{x}\right)$,
defined as \begin{align}
\delta\rho_{s}\left(\mathbf{x}\right) & \equiv-\frac{1}{\pi}\nabla\phi\left(\mathbf{x}\right),\label{eq:field_phi_def}\end{align}
we obtain the equivalent description%
{}\begin{align*}
S_{\left(1\right)}^{\text{w}} & =\frac{1}{2\pi^{2}}\frac{1}{\beta L}\sum_{\mathbf{q}}\left|\phi\left(\mathbf{q}\right)\right|^{2}\Biggl\{\frac{\omega_{m}^{2}}{4}\frac{1}{\mathcal{D}_{0}}+\\
 & \left.+q^{2}\frac{1+\left[\chi_{0,s}\left(\mathbf{q}\right)+\chi_{0,n}\left(\mathbf{q}\right)\right]v\left(q,0\right)}{\chi_{0,s}\left(\mathbf{q}\right)\left[1+\chi_{0,n}\left(\mathbf{q}\right)v\left(q,0\right)\right]}\right\} .\end{align*}
In the regime of Eq. (\ref{eq:condition_dynamical_screening}), we
can approximate the action by \begin{align*}
S_{\left(1\right)}^{\text{w}} & \simeq\frac{1}{2\pi^{2}}\frac{1}{\beta L}\sum_{\mathbf{q}}\Biggl\{\frac{\omega_{m}^{2}}{4}\frac{1}{\mathcal{D}_{0}}+\\
 & \left.+q^{2}\frac{\left[\gamma\mathcal{N}_{s,1D}^{0}+2\mathcal{N}_{n,1D}^{0}\frac{Dq^{2}}{\left|\omega_{m}\right|}\right]v\left(q,0\right)}{2\gamma\mathcal{N}_{n,1D}^{0}\mathcal{N}_{s,1D}^{0}\frac{Dq^{2}}{\left|\omega_{m}\right|}v\left(q,0\right)}\right\} \left|\phi\left(\mathbf{q}\right)\right|^{2},\\
 & =\frac{1}{2\pi^{2}}\frac{1}{\beta L}\sum_{\mathbf{q}}\left\{ \frac{\left|\omega_{m}\right|}{2\mathcal{N}_{n,1D}^{0}D}+\frac{\omega_{m}^{2}}{4\mathcal{D}_{0}}+\frac{q^{2}}{\gamma\mathcal{N}_{s,1D}^{0}}\right\} \left|\phi\left(\mathbf{q}\right)\right|^{2},\end{align*}
which shows that the term $\sim q^{2}\left|\omega_{m}\right|$ in
Eq. (\ref{eq:S_diff_theta}) translates into a relevant term $\sim\left|\omega_{m}\right|$
(in the RG-sense) when expressed in terms of $\phi\left(\mathbf{q}\right)$.
Another way to see this detrimental effect is through the order-parameter
correlation function $F\left(\mathbf{x}\right)=\Delta_{0}^{2}e^{-\frac{1}{2}\left\langle T_{\tau}\left[\theta\left(\mathbf{x}\right)-\theta\left(0\right)\right]^{2}\right\rangle }$
{[}cf. Eq. (\ref{eq:sc_correlation_function})], which vanishes due
to the infrared divergence of the phase-correlator $\left\langle T_{\tau}\left[\theta\left(\mathbf{x}\right)-\theta\left(0\right)\right]^{2}\right\rangle \equiv\frac{2}{\beta L}\sum_{\mathbf{q}}\frac{1-\cos\mathbf{q}.\mathbf{x}}{2\mathcal{N}_{n}^{0}Dq^{2}\left|\omega_{m}\right|+\mathcal{D}_{0}q^{2}}\rightarrow\infty$.
%
{}%
{}%
{}%
{}%
{} %
{}%
{}

Only at high-frequencies $\frac{2e^{2}}{\epsilon_{\textrm{r}}}\mathcal{N}_{n}^{0}Dq^{2}\ln\frac{2}{qr_{0}}\ll\left|\omega_{m}\right|$
(cf. dark gray area in Fig. \ref{fig:screening_regime_1D}), and provided
Eq. (\ref{eq:xi0n}) is still valid, or in the limit of very low electronic
density of states in the 1DEG, we recover the action of Eq. (\ref{eq:S_isolated_phase_only})
describing again unscreened plasma modes. Physically, at such high
frequencies the response $\chi_{0,n}\left(\mathbf{q}\right)$ of the
1DEG vanishes and the superconducting wire is effectively unscreened.

\begin{widetext}

\begin{table}[h]
\begin{tabular}{|c|c|c|c|c|c|c|c|c|c|}
\hline
$r_{0}\simeq d$  & $L$ & $\mathcal{D}_{0}$ & $\mathcal{N}_{s,1D}^{0}\simeq\mathcal{N}_{n,1D}^{0}$ & $\Delta_{0}$ & $D$  & $\epsilon_{\text{r}}$ & $w_{\text{film}}$ & $k_{\text{TF}}^{2D}$ & $\mathcal{N}_{n,2D}^{0}$\tabularnewline
\hline
\hline
10 nm & 100 $\mu\text{m}$  & 8.6 $10^{35}\:\frac{1}{\text{kg.m}}$ & $10^{29}\frac{1}{\text{m}.\text{J}}$ & 1 K & 0.01 $\frac{\text{m}^{2}}{\text{s}}$  & 1 & 100 nm & 1 $\text{nm}^{-1}$ & $10^{38}\frac{1}{\text{m}^{2}.\text{J}}$\tabularnewline
\hline
\end{tabular}

\caption{\label{tab:parameters}Parameters used in the calculations. Order-of-magnitude
estimations of $r_{0}$ and $L$ have been extracted from experiments
on superconducting aluminium wires with coherence length estimated
as $\xi_{0}\sim100$ nm {[}cf. Ref. \onlinecite{altomare06_experimental_evidence_of_qps}]. }

\end{table}

\end{widetext}

%
{}

\subsection{Screening by a diffusive metallic film}

Now we focus our attention on the system of Fig. \ref{fig:system}(b),
which represents a superconducting wire coupled to a normal diffusive
film of width $w_{\text{film}}$. In this case, the effective action
Eq. (\ref{eq:S_eff}) writes in Fourier space

\begin{align}
S_{\left(2\right)}^{\textrm{w}} & \simeq\frac{1}{2\beta L}\sum_{\mathbf{q}}\Biggl\{\omega_{m}\theta^{*}\left(\mathbf{q}\right)\rho_{s}\left(\mathbf{q}\right)+q^{2}\mathcal{D}\left(\mathbf{q}\right)\left|\theta\left(\mathbf{q}\right)\right|^{2}+\nonumber \\
 & +\left[\chi_{0,s}^{-1}\left(\mathbf{q}\right)+v\left(q,0\right)\right]\left|\rho_{s}\left(\mathbf{q}\right)\right|^{2}\Biggr\}+\nonumber \\
 & +\frac{1}{2\beta LL_{\perp}}\sum_{\mathbf{q},k}\Biggl\{\left[\chi_{0,n}^{-1}\left(\mathbf{q},k\right)+v\left(q,k,0\right)\right]\left|\rho_{n}\left(\mathbf{q},k\right)\right|^{2}+\nonumber \\
 & +v\left(q,k,d\right)\left[\rho_{s}^{*}\left(\mathbf{q}\right)\rho_{n}\left(\mathbf{q},k\right)+\rho_{s}\left(\mathbf{q}\right)\rho_{n}^{*}\left(\mathbf{q},k\right)\right]\Biggr\},\label{eq:S_eff_2D_d0}\end{align}
where the Coulomb interaction is {[}compare to Eq. (\ref{eq:v_q_z})]\begin{align*}
v\left(q,k,d\right) & =\frac{2\pi e^{2}}{\epsilon_{\text{r}}}\frac{e^{-\sqrt{q^{2}+k^{2}}d}}{\sqrt{q^{2}+k^{2}}}.\end{align*}
In this case, the presence of the superconducting wire breaks the
translational symmetry of the system in the direction perpendicular
to the wire. Consequently, the perpendicular momentum $k$ in the
plane is not conserved and the Coulomb interaction couples the density
modes in the wire with momentum $q$ to all the modes in the plane
with momentum $k$. As before, in order to obtain an effective model
for the phase field, we must integrate over the density fields $\rho_{n}\left(\mathbf{q},k\right)$
and $\rho_{s}\left(\mathbf{q}\right)$, which yields%
{}\begin{align}
S_{\left(2\right)}^{\textrm{w}} & \simeq\frac{1}{2}\frac{1}{\beta L}\sum_{\mathbf{q}}\Biggl\{\frac{\omega_{m}^{2}}{4}\frac{\chi_{0,s}\left(\mathbf{q}\right)}{1+\chi_{0,s}\left(\mathbf{q}\right)\left[v\left(q,0\right)-v_{\text{eff}}\left(\mathbf{q}\right)\right]}+\nonumber \\
 & +q^{2}\mathcal{D}\left(\mathbf{q}\right)\Biggr\}\left|\theta\left(\mathbf{q}\right)\right|^{2},\label{eq:S_eff_2D_phase_only}\end{align}
%
{}where $v_{\text{eff}}\left(\mathbf{q}\right)$ is an effective 1D-potential
encoding all the screening properties of the diffusive film \begin{align}
v_{\text{eff}}\left(\mathbf{q}\right) & \equiv\frac{1}{L_{\perp}}\sum_{k}\frac{v\left(q,k,d\right)^{2}\chi_{0,n}\left(\mathbf{q},k\right)}{1+v\left(q,k,0\right)\chi_{0,n}\left(\mathbf{q},k\right)}.\label{eq:v_eff}\end{align}
%
{}In this case, the susceptibility at low energies writes\citep{akkermans}\begin{align}
\chi_{0,n}\left(\mathbf{q},k\right) & \simeq2\mathcal{N}_{n,2D}^{0}\frac{D\left(q^{2}+k^{2}\right)}{D\left(q^{2}+k^{2}\right)+\left|\omega_{m}\right|},\label{eq:xi0n_2D}\end{align}
where $\mathcal{N}_{n,2D}^{0}$ is the 2D density of states at the
Fermi energy in the normal metal. Here again, we neglect Anderson-localization
effects in the metal by assuming that the length of the wire is $L\ll\xi_{\textrm{film}}$,
where $\xi_{\textrm{film}}$ is the localization length in the film.

In what follows, we derive analytical expressions for the model in
the limiting cases of static and dynamic regimes.

\subsubsection{Static screening limit $\left|\omega_{m}\right|\ll Dq^{2}$}

This region corresponds to the ligh gray area in Fig. (\ref{fig:screening_regime_2D}).
A relevant length scale which naturally appears is the 2D Thomas-Fermi
screening length, $\lambda_{\text{TF}}^{\textrm{2D}}=\frac{\epsilon_{\text{r}}}{2e^{2}\mathcal{N}_{n,2D}^{0}}$,
beyond which the long-range Coulomb potential is completely screened.
This quantity defines the Thomas-Fermi wavevector $k_{\text{TF}}=\frac{4\pi e^{2}\mathcal{N}_{n,2D}^{0}}{\epsilon_{\text{r}}}$.%
{}%
{} In the experimentally relevant limit $k_{\text{TF}}d\gg1$, the effective
potential $v_{\text{eff}}\left(\mathbf{q}\right)$ reduces to %
{}\begin{align}
v_{\text{eff}}\left(\mathbf{q}\right) & \simeq\frac{2e^{2}}{\epsilon_{\text{r}}}\left[K_{0}\left(2qd\right)-\frac{\pi}{2}\frac{\left|\omega_{m}\right|}{Dqk_{\text{TF}}}\right].\label{eq:v_eff_static_limit}\end{align}
The first and second term in the above expression are consistent with
the static and dissipative contributions, respectively, to the effective
screened interaction obtained for a Tomonaga-Luttinger liquid electrostatically
coupled to a diffusive 2DEG (cf. Ref. \onlinecite{cazalilla06_dissipative_transition}).
%
{} The static screening provided by the 2DEG {[}first term in Eq. (\ref{eq:v_eff_static_limit})]
cuts the logarithmic divergence of the bare intrawire Coulomb interaction
$v\left(q,0\right)$. The relationship between the second term in
Eq. (\ref{eq:v_eff_static_limit}) and the dissipative contribution
$\sim q\left|\omega_{m}\right|$ in Ref. \onlinecite{cazalilla06_dissipative_transition}
can be made explicit with the introduction of the field $\phi\left(\mathbf{x}\right)$,
defined in Eq. (\ref{eq:field_phi_def}).

In the limit $qd\rightarrow0$ (with $\left|\omega_{m}\right|\ll Dq^{2}$),
the effective potential Eq. (\ref{eq:v_eff_static_limit}) can be
further simplified to $v_{\text{eff}}\left(\mathbf{q}\right)\simeq\frac{2e^{2}}{\epsilon_{\text{r}}}\left[\ln\left(\frac{1}{qd}\right)-\frac{\pi}{2}\frac{\left|\omega_{m}\right|}{Dqk_{\text{TF}}}\right],$
and when replaced in Eq. (\ref{eq:S_eff_2D_phase_only}) yields\begin{align*}
S_{\left(2\right)}^{\textrm{w}} & \simeq\frac{1}{2\beta L}\sum_{\mathbf{q}}\Bigl[\frac{\omega_{m}^{2}}{4}\chi_{s}\left(0\right)-\frac{\pi e^{2}\chi_{s}^{2}\left(0\right)\left|\omega_{m}\right|^{3}}{Dk_{\text{TF}}q}+q^{2}\mathcal{D}_{0}\Bigr]\left|\theta\left(\mathbf{q}\right)\right|^{2},\end{align*}
{}%
{}where $\chi_{s}\left(0\right)\simeq\chi_{0,s}\left(0\right)\left[1+\frac{2e^{2}}{\epsilon_{\text{r}}}\chi_{0,s}\left(0\right)\ln\left(2d/r_{0}\right)\right]^{-1}$
is the RPA susceptibility of the wire.

Note that in the limit $D\rightarrow\infty$ (no dissipation in the
normal metallic film), we recover again the action of a Tomonaga-Luttinger
liquid with short-range interactions, with plasma modes obeying a
linear dipersion relation\citep{giamarchi_book_1d}. In the more general
case of a finite $D$, we have the relation dispersion

\begin{align}
\omega^{2}\left(q\right)\chi_{s}\left(0\right)+i\frac{e^{2}}{\epsilon_{\text{r}}}\frac{\pi\omega^{3}\left(q\right)\chi_{s}^{2}\left(0\right)}{Dqk_{\text{TF}}}+q^{2}\mathcal{D}_{0} & =0,\label{eq:dispersion_relation_2D_static}\end{align}
which describes plasma modes with approximately linear dispersion
relation, and with a width $\Gamma\left(q\right)=-\text{Im}\left[\omega\left(q\right)\right]\sim\frac{\pi e^{2}\mathcal{D}_{0}}{2\epsilon_{\text{r}}Dk_{\text{TF}}\chi_{s,0}\left(0\right)}q$.
Note that this result only applies in the limit $\left|\omega\left(q\right)\right|\ll\left\{ Dq^{2},\frac{Dqk_{\text{TF}}\epsilon_{\text{r}}}{\pi e^{2}\chi_{s}\left(0\right)}\right\} $,
and eventually breaks down in the limit $q\rightarrow0$, meaning
that this is not the relevant regime at low energies.

\subsubsection{Dynamical screening limit $\left|\omega_{m}\right|\gg Dq^{2}$}

\begin{figure}[h]
\includegraphics[bb=0bp 0bp 351bp 210bp,clip,scale=0.7]{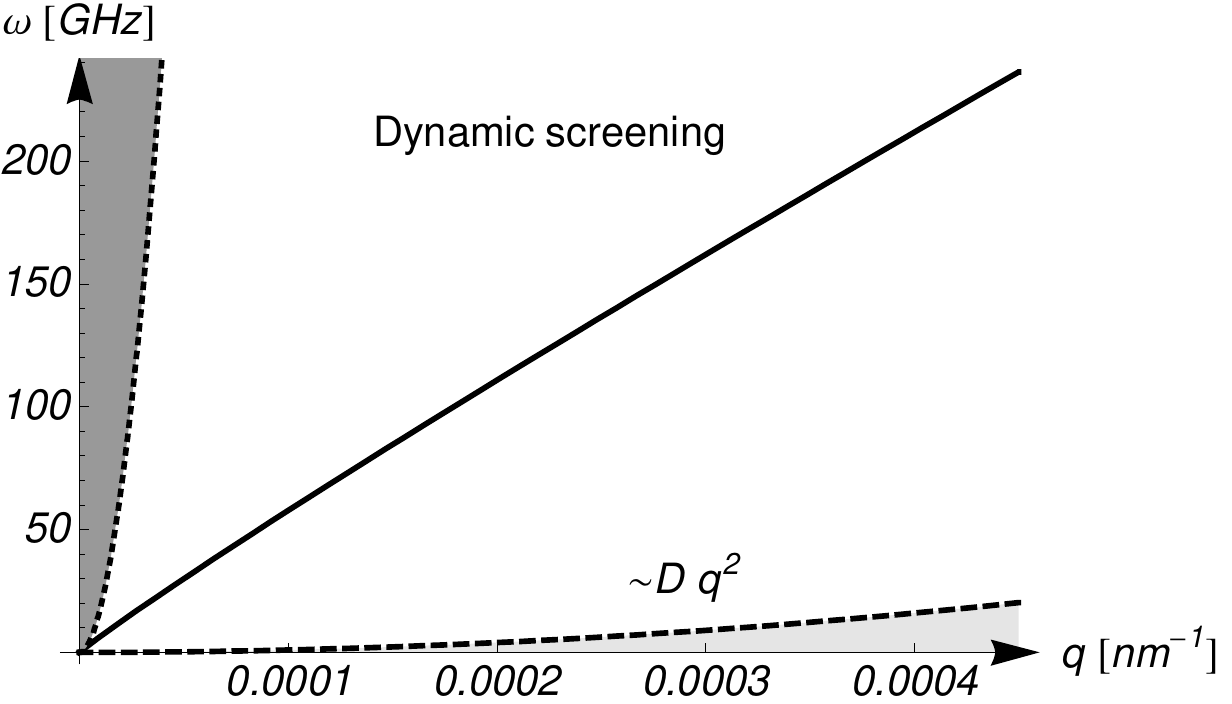}

\caption{\label{fig:screening_regime_2D}Screening regimes for a superconducting
wire capacitively coupled to a diffusive 2DEG. The dispersion relation
(solid line) results from Eq. (\ref{eq:dispersion_relation_2D}),
valid in the regime $Dq^{2}\ll\left|\omega_{m}\right|\ll\frac{Dk_{\text{TF}}}{d}$.
As in Fig. \ref{fig:screening_regime_1D}, the dashed line $\omega=Dq^{2}$
separates the regime of static (light gray area) from that of dynamic
(white area) screening. For the same parameters as in Fig. \ref{fig:screening_regime_1D},
the plasma mode crosses over between the two regimes at higher frequencies.
The unscreened region corresponds to the dark gray area.}

\end{figure}

\begin{figure}[h]
\includegraphics[bb=0bp 0bp 350bp 190bp,clip,scale=0.7]{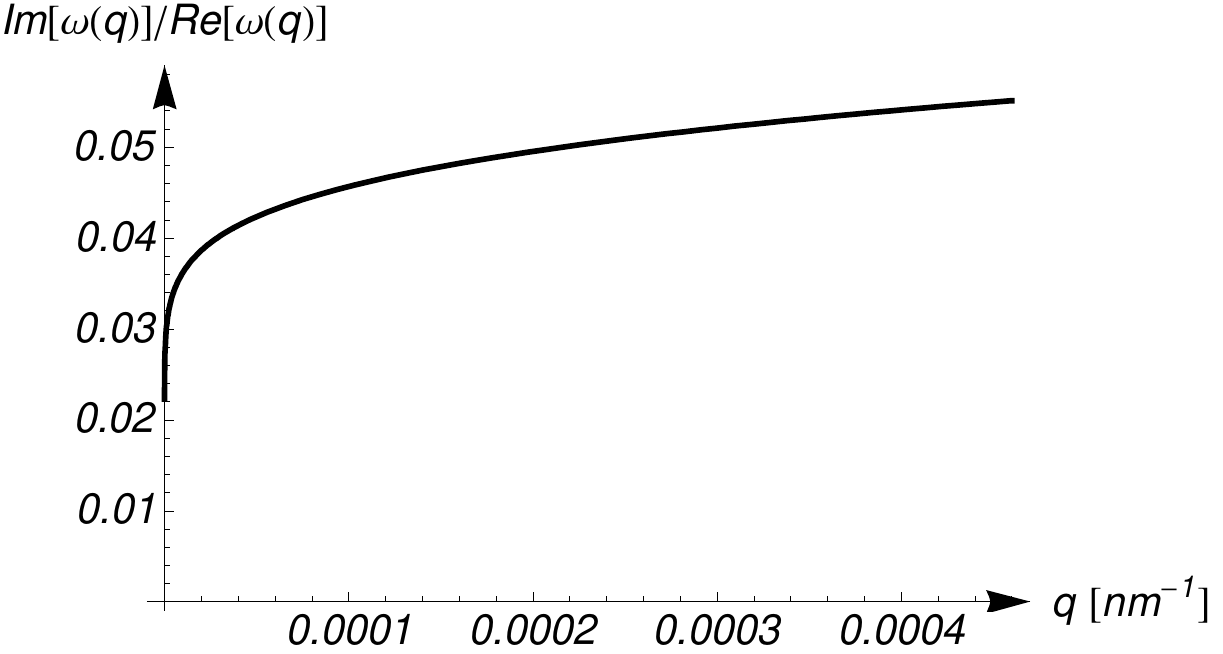}

\caption{\label{fig:im_real_ratio}Ratio $\frac{\text{Im}\left[\omega\left(q\right)\right]}{\text{Re}\left[\omega\left(q\right)\right]}$vs
$q$ for typical experimental parameters (cf. Table \ref{tab:parameters}).
The plasma mode becomes better defined as $q\rightarrow0$. }

\end{figure}

With realistic estimates for the experimental parameters (cf. Table
\ref{tab:parameters}), the regime $\left|\omega_{m}\right|\gg Dq^{2}$
is the most relevant in practical realizations. In this regime (white
area in Fig. \ref{fig:screening_regime_2D}), the effective potential
$v_{\text{eff}}\left(\mathbf{q}\right)$ in Eq. (\ref{eq:v_eff})
can be approximated as%
{} %
{}\begin{align*}
v_{\text{eff}}\left(\mathbf{q}\right) & \approx\frac{2e^{2}}{\epsilon_{\text{r}}}\left[f\left(2k_{\text{TF}}d\right)-f\left(\frac{2\left|\omega_{m}\right|d}{Dk_{\text{TF}}}\right)\right]\end{align*}
where we have defined $f\left(z\right)\equiv-e^{z}\text{Ei}\left(-z\right)$,
with $\text{Ei}\left(x\right)$ the exponential integral function\citep{abramowitz_math_functions}.
If the additional condition $\left|\omega_{m}\right|\ll\frac{Dk_{\text{TF}}}{d}$
holds, the effective potential can be further simplified to \begin{align*}
v_{\text{eff}}\left(\mathbf{q}\right) & \simeq\frac{2e^{2}}{\epsilon_{\text{r}}}\ln\left(\frac{2\left|\omega_{m}\right|}{Dk_{\text{TF}}}d\right).\end{align*}
Using this expression and Eq. (\ref{eq:v_q_z}), the phase-only action
of Eq. (\ref{eq:S_eff_2D_phase_only}) writes %
{}\begin{align*}
S_{\left(2\right)}^{\textrm{w}} & \simeq\frac{1}{2\beta L}\sum_{\mathbf{q}}\Biggl\{\frac{\omega_{m}^{2}}{4}\frac{\chi_{0,s}\left(0\right)}{1+\frac{2e^{2}}{\epsilon_{\text{r}}}\chi_{0,s}\left(0\right)\ln\left(\frac{Dk_{\text{TF}}}{dr_{0}q\left|\omega_{m}\right|}\right)}+\\
 & +q^{2}\mathcal{D}_{0}\Biggr\}\left|\theta\left(\mathbf{q}\right)\right|^{2},\end{align*}
resulting in the equation of motion (in the limit $\mathbf{q}\rightarrow0$)\begin{align}
-\frac{\omega^{2}}{4}+q^{2}\frac{2e^{2}}{\epsilon_{\text{r}}}\mathcal{D}_{0}\left(\ln\left|\frac{Dk_{\text{TF}}}{dr_{0}q\omega}\right|+i\frac{\pi}{2}\right) & =0.\label{eq:dispersion_relation_2D}\end{align}
 %
{} %
In Fig. \ref{fig:screening_regime_2D} we show (solid line) the dispersion
relation obtained from Eq. (\ref{eq:dispersion_relation_2D}) (i.e.,
real component of $\omega\left(q\right)$). Contrarily to the case
studied in Sec. \ref{sub:screening1D}, the resulting plasma mode
is not damped in the limit $q\rightarrow0$, i.e., a dispersive real
component survives. %
{}In order to investigate the dynamics of the phase-field at low-energies,
we study the ratio $\text{Im}\left[\omega\left(q\right)\right]/\text{Re}\left[\omega\left(q\right)\right]$
vs. $q$ (cf. Fig. \ref{fig:im_real_ratio}). From Eq. (\ref{eq:dispersion_relation_2D})
it is possible to show that in the limit $q\rightarrow0$%
{}\begin{align}
\frac{\text{Im}\left[\omega\left(q\right)\right]}{\text{Re}\left[\omega\left(q\right)\right]} & \simeq\frac{\pi}{4}\left[\ln\left(\frac{Dk_{\text{TF}}}{q^{2}dr_{0}\sqrt{\frac{8e^{2}}{\epsilon_{\text{r}}}\mathcal{D}_{0}}}\right)\right]^{-1},\label{eq:im_real_ratio_2D}\end{align}
meaning that the width of the plasma mode decreases at low energies,
resulting in a well-defined excitation. Note the difference with respect
to the screening provided by a 1DEG, where the damping of the plasmon
was complete in the limit $q\rightarrow0$. The origin of this difference
lies in the additional degree of freedom $k$ (momentum perpendicular
to the wire), which smears (upon integration) the dependence on the
damping factor $\left|\omega_{m}\right|$ in the susceptibility {[}cf.
Eq. (\ref{eq:xi0n_2D})] .

%
{}%
{}

In the regime of frequencies $\frac{Dk_{\text{TF}}}{d}\ll\left|\omega_{m}\right|\ll Dk_{\text{TF}}^{2}$,
the effective potential $v_{\text{eff}}\left(\mathbf{q}\right)$ writes
\begin{align*}
v_{\text{eff}}\left(\mathbf{q}\right) & \approx\frac{2e^{2}}{\epsilon_{\text{r}}}\frac{1}{2k_{\textrm{TF}}d}\left[1-\frac{Dk_{\text{TF}}^{2}}{\left|\omega_{m}\right|}\right].\end{align*}
and the phase only action is\begin{align*}
S_{\left(2\right)}^{\textrm{w}} & \simeq\frac{1}{2}\frac{1}{\beta L}\sum_{\mathbf{q}}\Biggl\{\frac{\omega_{m}^{2}}{4}\frac{\chi_{0,s}\left(0\right)}{1+\frac{2e^{2}}{\epsilon_{\text{r}}}\chi_{0,s}\left(0\right)\left[\ln\left(\frac{2}{qr_{0}}\right)+\frac{Dk_{\text{TF}}}{2d\left|\omega_{m}\right|}\right]}+\\
 & +q^{2}\mathcal{D}_{0}\Biggr\}\left|\theta\left(\mathbf{q}\right)\right|^{2}.\end{align*}
 In this limit, the equation of motion of the field $\theta\left(\mathbf{q}\right)$
is%
{} \begin{align*}
-\omega^{2}\left(q\right)+q^{2}\mathcal{D}_{0}\frac{8e^{2}}{\epsilon_{\text{r}}}\left[\ln\left(\frac{2}{qr_{0}}\right)+\frac{iDk_{\text{TF}}}{2d\omega\left(q\right)}\right] & =0.\end{align*}
Note that in this regime, the dissipative effects are even weaker
and the dispersion relation ressembles that of the (unscreened) Mooij-Sch{\"o}n
mode. Eventually in the limit $\left|\omega_{m}\right|\gg Dk_{\text{TF}}^{2}$,
the response $\chi_{0,n}\left(\mathbf{q}\right)$ of the 2DEG vanishes
and the wire is effectively in the unscreened regime where the Mooij-Sch{\"o}n
plasma mode of Eq. (\ref{eq:dispersion_relation_mooij_schoen}) is
fully recovered.

\section{\label{sec:dynamic_conductivity}Dissipative effects in the dynamic
conductivity}

In this section we study the consequences of thedissipative effects
on the dynamic conductivity of the wire $\sigma\left(q,\omega\right)$,
i.e., the ratio between the current density and the local electric
field $j\left(q,\omega\right)=\sigma\left(q,\omega\right)E\left(q,\omega\right)$.
This quantity is of interest because its real part $\text{Re}\left[\sigma\left(q,\omega\right)\right]$
provides information on the dissipation and absorption properties,
which result in our case from the friction mediated by the Coulomb
interaction\citep{note_technical_difficuly_dynamic_conductivity}.

\begin{figure}[h]
\includegraphics[bb=0bp 0bp 660bp 400bp,clip,scale=0.45]{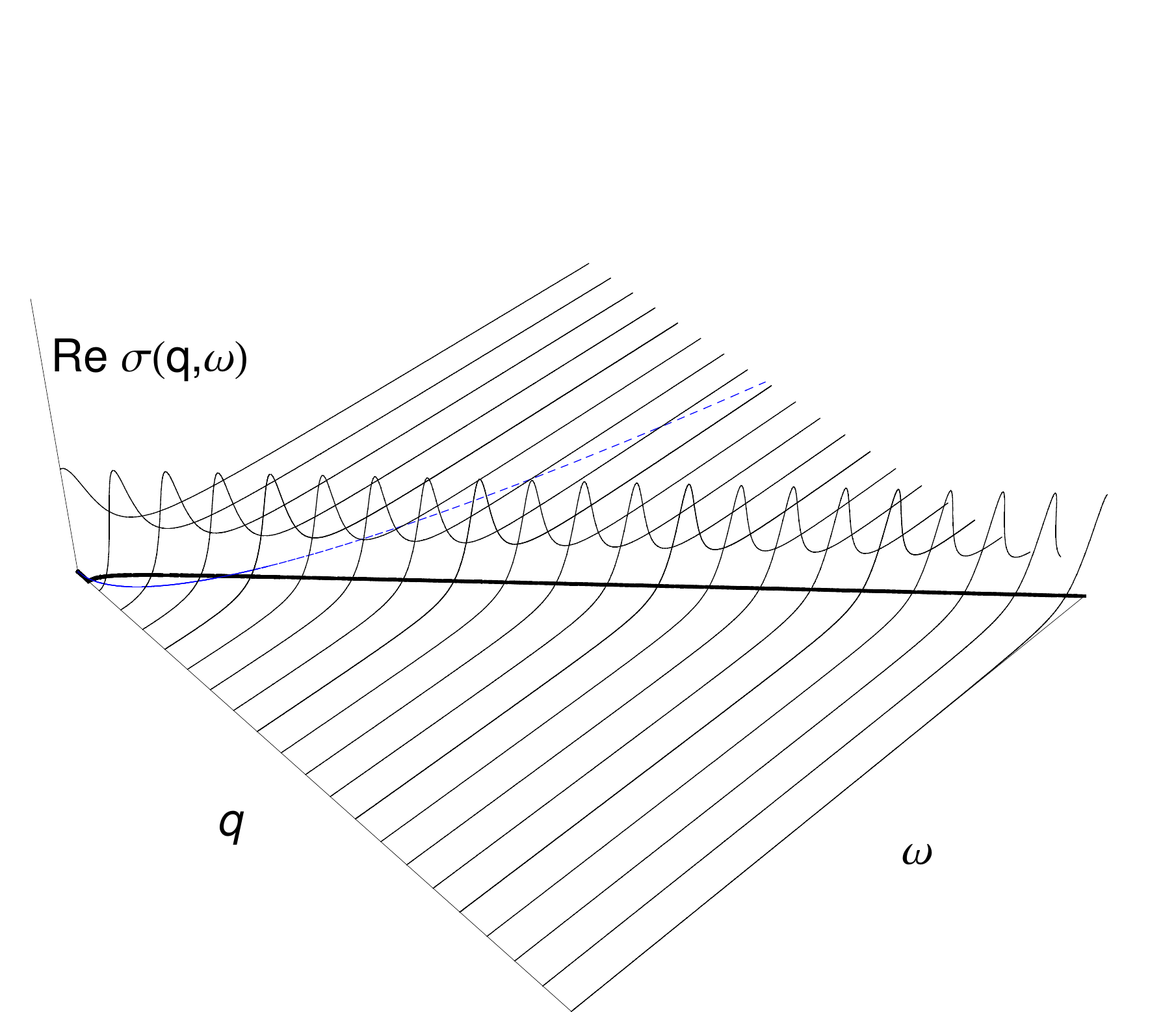}

\caption{\label{fig:dynamic_conductivity_1D}Dynamic conductivity $\text{Re }\sigma\left(q,\omega\right)$
of a superconducting wire dynamically screened by a diffusive 1DEG.
The plasma mode, which is better defined at high energy and momentum
becomes completely damped in the limit $\left\{ \omega,q\right\} \rightarrow0$
by the effects of the dissipative environment. }

\end{figure}

The response of the system to an external electromagnetic field can
be obtained by the means of the minimal coupling $-i\nabla\rightarrow-i\nabla-\frac{e}{c}A$
in the microscopic Hamiltonian Eq. (\ref{eq:h_s})%
{}. For a superconducting wire at $T=0$ and in absence of quasiparticle
excitations, the total current density is given by\begin{align*}
j\left(\mathbf{x}\right) & =j_{p}\left(\mathbf{x}\right)+j_{d}\left(\mathbf{x}\right),\\
j_{p}\left(\mathbf{x}\right) & =\frac{2e}{c}\mathcal{D}_{0}\nabla\theta\left(\mathbf{x}\right),\\
j_{d}\left(\mathbf{x}\right) & =-\left(\frac{2e}{c}\right)^{2}\mathcal{D}_{0}A\left(\mathbf{x}\right),\end{align*}
where $j_{p}$ and $j_{d}$ are, respectively, the paramagnetic and
diamagnetic contributions to the current density. The linear response
to an applied electromagnetic field is given by the current-current
susceptibility of the wire \begin{align*}
\chi_{jj}\left(\mathbf{q}\right) & =\left.\frac{\delta\ln Z}{\delta A_{\mathbf{q}}\delta A_{-\mathbf{q}}}\right|_{A=0}=\left\langle j_{p}\left(\mathbf{q}\right)j_{p}\left(-\mathbf{q}\right)\right\rangle -\mathcal{D}_{0}\left(\frac{2e}{c}\right)^{2}.\end{align*}
%
{} The conductivity is in turn related to the current-current susceptibility
by the relation\citep{mahan2000,giamarchi_book_1d} $\sigma\left(\mathbf{q}\right)=-\frac{\chi_{jj}\left(\mathbf{q}\right)}{\omega_{m}}$,
upon analytical continuation to real frequencies $\left.\sigma\left(\mathbf{q}\right)\right|_{i\omega_{m}\rightarrow\omega+i\delta}=\sigma\left(q,\omega\right)$.%
{} In terms of the phase field $\theta\left(\mathbf{q}\right)$, the
conductivity reads \begin{align}
\sigma\left(\mathbf{q}\right) & \equiv-\left(\frac{2e}{c}\right)^{2}\left[-\frac{\mathcal{D}_{0}}{\omega_{m}}+\frac{\mathcal{D}_{0}^{2}q^{2}}{\omega_{m}}\left\langle \theta\left(\mathbf{q}\right)\theta\left(-\mathbf{q}\right)\right\rangle \right].\label{eq:phase_dynamic_conductivity}\end{align}

Let us first study the response of an ideally isolated wire (cf. Sec.
\ref{sub:unscreened}) to the electromagnetic field. At $T=0$ we
obtain%
{} \begin{align}
\text{Re}\left[\sigma\left(q,\omega\right)\right] & =\frac{\pi}{2}\mathcal{D}_{0}\left(\frac{2e}{c}\right)^{2}\delta\left(\omega-\omega\left(q\right)\right),\label{eq:real_conductivity_unscreened}\end{align}
where $\omega\left(q\right)=\sqrt{\frac{8e^{2}}{\epsilon_{\text{r}}}\ln\left(\frac{2}{\left|qr_{0}\right|}\right)q^{2}\mathcal{D}_{0}}$
is the energy of the Mooij-Sch{\"o}n plasmon {[}cf. Eq. (\ref{eq:dispersion_relation_mooij_schoen})].
The real part of the conductivity tells us that the system absorbs
energy at the frequency $\omega=\omega\left(q\right)$, which in this
case are well-defined excitations (i.e., delta-functions). Note that
in the limit $q\rightarrow0$, Eq. (\ref{eq:real_conductivity_unscreened})
allows to recover the Drude peak at $\omega=0$, which is expected
for a superconductor\citep{giamarchi_book_1d,mahan2000}.

\begin{figure}[h]
\includegraphics[bb=0bp 0bp 250bp 140bp,clip]{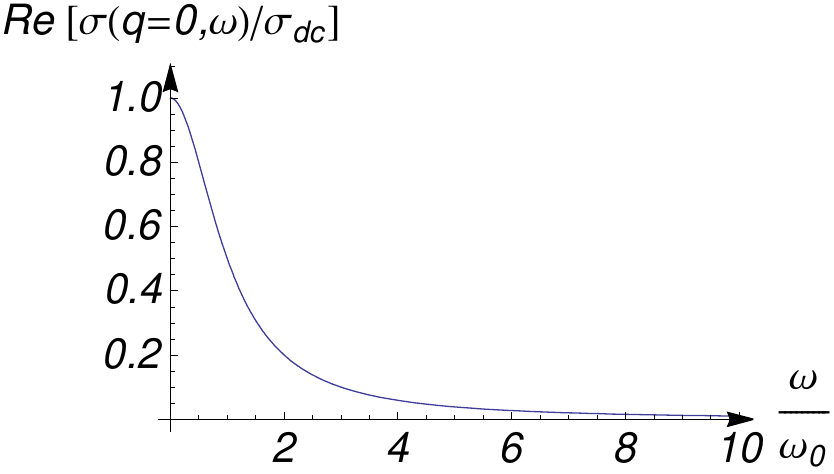}

\caption{\label{fig:conductivity_1D_q0}dc-conductivity $\sigma\left(\omega\right)=\text{Re}\left[\sigma\left(q=0,\omega\right)\right]$
of a superconducting wire screened by a diffusive 1DEG. The values
on the axis are normlized to $\sigma_{\textrm{dc}}=\sigma\left(\omega=0\right)=\left(\frac{2e}{c}\right)^{2}2D\mathcal{N}_{n,1D}^{0}$
and $\omega_{0}\equiv\frac{\mathcal{D}_{0}}{2\mathcal{N}_{n,1D}^{0}D}$.
The plasma mode at finite $q$, is completely damped in the limit
$q=0$ (cf. Fig. \ref{fig:dynamic_conductivity_1D}). }

\end{figure}

Let us now consider the case of a wire in the proximity to an electron
gas. We first study the case of screening by a diffusive 1DEG (cf.
Sec. \ref{sub:1D_screening_dynamic}), where the effects of dissipation
are at their strongest. Using the action of Eq. (\ref{eq:S_eff_1D_phase_only})
to evaluate the formula of the conductivity {[}cf. Eq. (\ref{eq:phase_dynamic_conductivity})]
we obtain the expression (valid at $T=0$)%
{} \begin{align}
\text{Re}\left[\sigma\left(q,\omega\right)\right] & =\mathcal{D}_{0}^{2}\left(\frac{2e}{c}\right)^{2}\frac{q^{2}}{\omega}\text{Im}\Biggl[q^{2}\mathcal{D}_{0}-\frac{\left(\omega+i0^{+}\right)^{2}}{4}\times\nonumber \\
 & \times\frac{\chi_{0,s}\left(0\right)\left[1+\chi_{0,n}^{\textrm{ret}}\left(q,\omega\right)v\left(q,0\right)\right]}{1+\left[\chi_{0,s}\left(0\right)+\chi_{0,n}^{\textrm{ret}}\left(q,\omega\right)\right]v\left(q,0\right)}\Biggr]^{-1},\label{eq:real_conductivity_1D}\end{align}
where $\chi_{0,n}^{\textrm{ret}}\left(q,\omega\right)\equiv\lim_{\delta\rightarrow0^{+}}\left[\chi_{0,n}\left(\mathbf{q}\right)\right]_{i\omega_{m}\rightarrow\omega+i\delta}$
is the (disorder-averaged) retarded density-density correlation function
in the 1DEG. In Fig. \ref{fig:dynamic_conductivity_1D} we show the
result for $\text{Re}\left[\sigma\left(q,\omega\right)\right]$ of
Eq. (\ref{eq:real_conductivity_1D}) in the plane $q-\omega$. The
dispersion relation $\text{Re}\left[\omega\left(q\right)\right]$
vs. $q$ (thick solid line in the bottom plane) was calculated numerically
from Eq. (\ref{eq:S_eff_1D_phase_only}) and corresponds to the same
plot of Fig. \ref{fig:screening_regime_1D}. As mentioned before,
the absorption peaks of $\text{Re}\left[\sigma\left(q,\omega\right)\right]$
are centered at the frequency $\text{Re}\left[\omega\left(q\right)\right]$
of the plasma mode. The curve $Dq^{2}$ (blue dotted line) is plotted
in the bottom $q-\omega$ plane to visualize the different screening
regimes. Note that the dissipative effects in the normal wire (encoded
in a finite value of the diffussion constant $D$) are manifested
in this figure through the finite width $\Gamma\left(q\right)\simeq-\text{Im}\left[\omega\left(q\right)\right]$
of the plasmon peaks. Note in addition that the constant width $\Gamma\left(q\right)$
in the regime $\left|\omega\left(q\right)\right|\gg Dq^{2}$ is consistent
with the result for $\text{Im}\left[\omega\left(q\right)\right]$
of Fig. \ref{fig:im_real_1D}.

As $q\rightarrow0$, the plasmon peak merges smoothly into the dc-conductivity
value $\sigma_{\textrm{dc}}=\left(\frac{2e}{c}\right)^{2}2D\mathcal{N}_{n,\textrm{1D}}^{0}$,
which exactly corresponds to the dc-conductivity of the 1DEG (cf.
Fig. \ref{fig:conductivity_1D_q0})%
{}. This expression is obtained by replacing the expression of the action
of Eq. (\ref{eq:S_diff_theta}) in the general expression of the conductivity
Eq. (\ref{eq:phase_dynamic_conductivity}). Physically, this means
that the Coulomb interaction produces friction in the superconductor
through the dissipation existing the 1DEG. It also indicates that
the original plasma mode is no longer a well-defined excitation of
the system, and that the electromagnetic environment have profound
consequences in the excitation spectrum of the 1D superconductor.

As we mentioned before, far from the BKT quantum critical point, phase
slips are an irrelevant perturbation (in the RG-sense). In the case
of Luttinger liquids with short-range interactions, the perturbative
effect of phase-slips generates a power-law resistivity $\varrho\sim T^{\nu}$,
with $\nu$ a positive exponent\citep{giamarchi_attract_1d}. Although
we have neglected the perturbative effect of topological excitations
in our formalism, the fact that a finite resistivity at $T=0$ appears
in the superconducting wire indicates that their effect in the conductivity
might be negligible as compared to those induced by dissipation in
the electron gas.

\begin{figure}[h]
\includegraphics[bb=0bp 50bp 480bp 450bp,clip,scale=0.45]{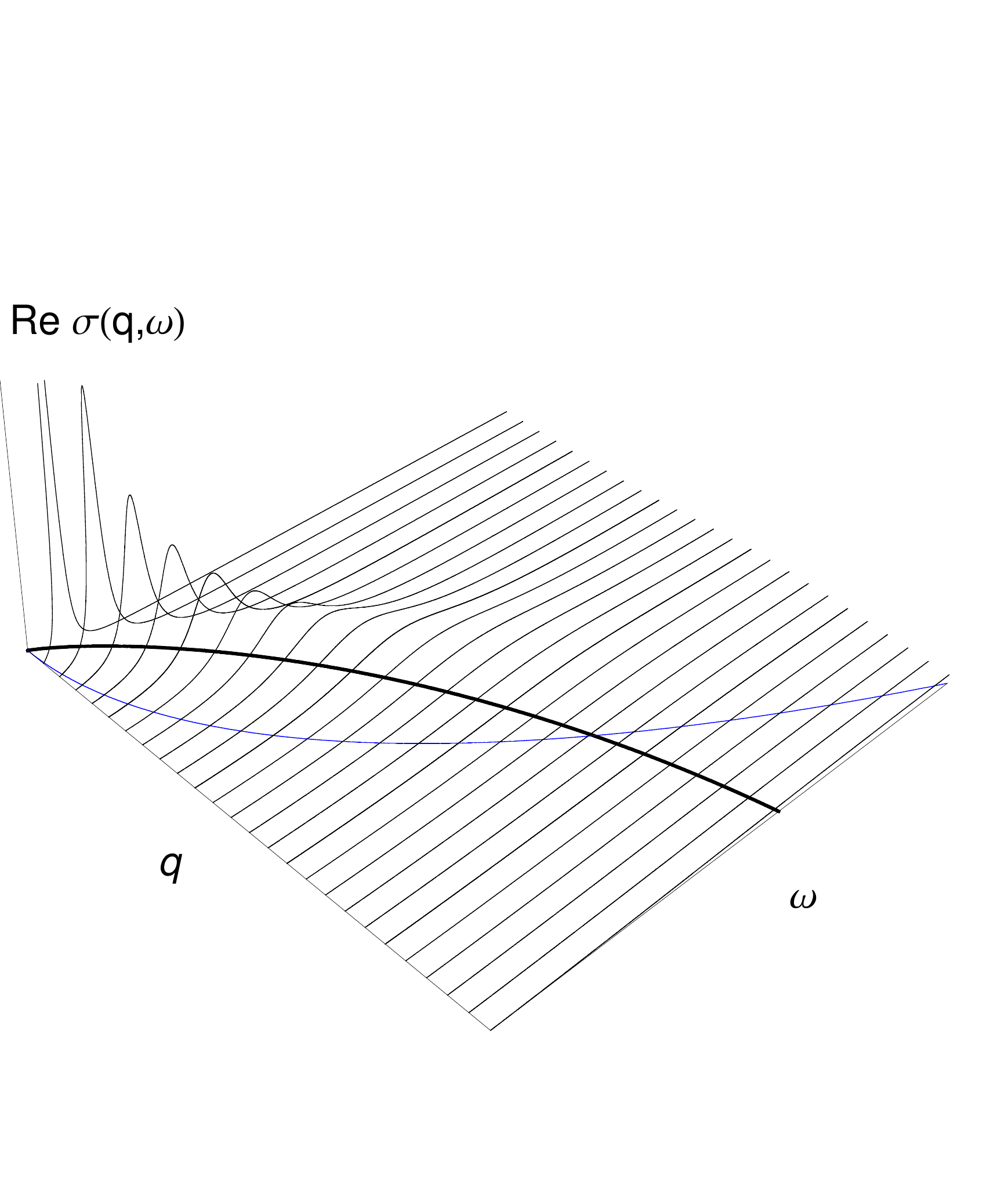}

\caption{\label{fig:dynamic_conductivity_2D}Dynamic conductivity $\text{Re}\left[\sigma\left(q,\omega\right)\right]$
of a superconducting wire dynamically screened by a diffusive 2DEG.
The plasma mode is completely damped at high energy and momentum,
but in the limit $\left\{ \omega,q\right\} \rightarrow0$ the effects
of dissipation vanish. }

\end{figure}
In the case of screening by a diffusive 2DEG, the expression of the
conductivity is given by the expression (valid at $T=0$)\begin{align*}
\text{Re}\left[\sigma\left(q,\omega\right)\right] & =\mathcal{D}_{0}^{2}\left(\frac{2e}{c}\right)^{2}\frac{q^{2}}{\omega}\text{Im}\biggl[q^{2}\mathcal{D}_{0}-\frac{\left(\omega+i0^{+}\right)^{2}}{4}\times\\
 & \times\frac{\chi_{0,s}\left(0\right)}{1+\chi_{0,s}\left(0\right)\left[v\left(q,0\right)-v_{\text{eff}}^{\textrm{ret}}\left(q,\omega\right)\right]}\biggr]^{-1},\end{align*}
%
{}where $v_{\text{eff}}^{\textrm{ret}}\left(q,\omega\right)\equiv\lim_{\delta\rightarrow0^{+}}\left[v_{\text{eff}}\left(\mathbf{q}\right)\right]_{i\omega_{m}\rightarrow\omega+i\delta}$.
Our main results in this case are presented in Fig. \ref{fig:dynamic_conductivity_2D}.
Contrarily to the case of Fig. \ref{fig:dynamic_conductivity_1D},
the plasmon peaks are worse defined at high energies, while at low
energies the width of the peak centered at $\text{Re}\left[\omega\left(q\right)\right]$
decreases and eventually vanishes in the limit $q\rightarrow0$, in
agreement with Eq. (\ref{eq:im_real_ratio_2D}) and Fig. (\ref{fig:im_real_ratio}).
Eventually, the plasmon peak merges into the superconducting Drude
peak at $\omega=0$.

The presence of an additional degree of freedom (i.e., momentum in
the plane perpendicular to the wire) is of central importance to understand
the vanishing of dissipation. Indeed, even in the dynamical screening
regime $Dq^{2}\ll\left|\omega_{m}\right|$ for which one would naively
think that dissipation effects are dominant, the presence of perpendicular
wavevectors $k$ satisfying the condition $\left|\omega_{m}\right|\ll Dk^{2}$
make the dissipative processes less important. Note in addition that
this condition is more easily satisfied in the limit $\left|\omega_{m}\right|\rightarrow0$.
These qualitative phase-space considerations allow to understand the
behavior of the effective 1D potential $v_{\text{eff}}\left(\mathbf{q}\right)$
of Eq. (\ref{eq:v_eff}), which produces a weaker (i.e., logarithmic)
dependence on the term $\left|\omega_{m}\right|$ encoding the dissipation.
The net result is that the 1D plasma modes are better defined in the
limit $q\rightarrow0$ and the frictional effects vanishes strictly
in the thermodynamical limit $L\rightarrow\infty$.

\section{\label{sec:Discussion}Discussion and Summary}

In this article we have studied the effects of the local electromagnetic
environment, provided by the presence of a non-interacting electron
gas, on the low-energy physics of a superconducting wire. In particular,
we have focused on the derivation of an effective phase-only action,
starting from the microscopic Hamiltonian of the system. We make extensive
use of the path-integral formalism, which enables to decouple the
superconducting and long-range Coulomb interactions by the means of
Hubbard-Stratonovich fields, and to expand the resulting action in
terms of quadratic deviations of these fields around the saddle-point
(i.e., Gaussian fluctuations). This treatment is equivalent to performing
the so-called RPA-approximation of the interacting problem\citep{mahan2000}.%
{} We have studied two cases in particular, namely, the screening provided
by a diffusive 1DEG, and a diffusive 2DEG placed at a distance $d\simeq r_{0}$
from the wire, with $r_{0}$ its radius. This would be the relevant
situation in practical realizations in, e.g., superconductor/normal
heterostructures made by the means of the ferroelectric field-effect
in Nb-doped SrTiO$_{3}$ layers\citep{Takahashi:2006rt}, or in electrically
controlled LaAlO$_{3}$/SrTiO$_{3}$ interfaces \citep{Reyren07_Superconducting_Interfaces_Between_Insulating_Oxides,Caviglia08_Electric_field_control_of_the_LaAlO3/SrTiO3_interface_ground_state}.

%
{}%
{}

%
{}

It is of interest to put our results in the context of other works
dealing with electrostatically coupled 1D systems. Among these, the
Coulomb drag effect\citep{rojo99_coulomb_drag_review}, where a finite
current $I_{1}$ is driven in one (the {}``active'') system, and
a finite voltage $V_{2}$ is induced in the other ({}``passive''
system),%
{} has received a great deal of attention both theoretically\citep{Duan93_supercurrent_drag_in_1D_conductors,Pustilnik_coulomb_drag,Flensberg98_Coulomb_Drag_of_Luttinger_Liquids_and_Quantum_Hall_Edges,Nazarov98_Current_Drag_in_Capacitively_Coupled_Luttinger_Constrictions,Klesse00_Coulomb_drag_between_quantum_wires}
and experimentally\citep{Giordano94_Cross-talk_effects_SIN_trilayers,Huang95_Observation_of_Supercurrent_Drag_between_Normal_Metal_and_Superconducting_Films,Farina04_Anomalous_drag_in_SN_films_by_radio_frequency_interference}.
Although closely related, the focus of our work is on the equilibrium
properties of the wire. %
{}%
{}

From the theoretical point of view, our work differs from the usual
Tomonaga-Luttinger liquid description of a purely 1D (i.e., one electronic
conduction channel) conductor, where the main mechanism of momentum
transfer is backscattering\citep{Flensberg98_Coulomb_Drag_of_Luttinger_Liquids_and_Quantum_Hall_Edges,Nazarov98_Current_Drag_in_Capacitively_Coupled_Luttinger_Constrictions,Klesse00_Coulomb_drag_between_quantum_wires,cazalilla06_dissipative_transition}.%
{}%
{} Indeed, it is worth to note that intra- and/or interwire backscattering
effects are absent in clean wires with a large number of electronic
channels, and this fact is correctly reproduced by our effective coarse-grained
theory {[}cf. Eq. (\ref{eq:S_eff})]. Therefore, in the language of
Tomonaga-Luttinger liquid physics, our treatment amounts to retaining
only forward scattering processes.

%
{}

Our results point towards a rich behavior of the 1D plasmon mode in
the wire, determined by the diffusive modes in the electron gas. Independently
of its dimensionality, in the static screening limit $\left|\omega_{m}\right|\ll Dq^{2}$,
the wire has a plasmon excitation which follows approximately a linear
dispersion relation. One could naively think that in that regime dissipative
effects are negligible. However, the complete solutions of Eqs. (\ref{eq:LL_dispersion_relation})
and (\ref{eq:dispersion_relation_2D_static}) indicate that this is
not the case. Indeed, we obtain sizable dissipative effects even in
the limit $\left|\omega\left(q\right)\right|\ll Dq^{2}$, which manifests
itself in the broadening of the 1D plasmon mode (cf. Figs. \ref{fig:dynamic_conductivity_1D}
and \ref{fig:dynamic_conductivity_2D}). Although technically challenging
from the experimental point of view, this broadening could be seen
in experiments of resonant inelastic Raman light-scattering\citep{goni_gas1d}
or in optical measurements of the dynamic conductivity or the reflection
coefficient\citep{buisson_dynamical_screening_thing_scwires}. %
{}

On the other hand, our results reveal that the dynamical screening
regime $Dq^{2}\ll\left|\omega_{m}\right|$ should be the most relevant
for experimental realizations (cf. Figs. \ref{fig:screening_regime_1D}
and \ref{fig:screening_regime_2D}). This is more or less evident
from the fact that the plasma mode essentially follows a linear dispersion
in the limit $q\rightarrow0$, while the boundary between the dynamical
and the static screening regimes (determined by the diffusive modes
in the electron gas) is $\sim Dq^{2}$. %
{}More importantly, in the limit $Dq^{2}\ll\left|\omega_{m}\right|$
the dimensionality of the electron gas is of central importance to
determine the low-energy properties of the wire. If the screening
is provided by a 1DEG, its dissipative processes are more efficiently
transferred to the superconducting wire in the limit $q\rightarrow0$.
As a consequence, the plasma mode becomes an ill-defined excitation
and the superconductor shows a finite dc-conductivity in the limit
$\omega=0$ (cf. Figs. \ref{fig:dynamic_conductivity_1D} and \ref{fig:conductivity_1D_q0}).
%
{}This effect could be seen, e.g., in dc-transport experiments on capacitively
coupled superconducting/normal wires systems%
{} (cf. Fig. \ref{fig:conductivity_1D_q0}).

When the screening is provided by a 2DEG, acoustic plasma modes with
a vanishing width are recovered in the limit $q\rightarrow0$, which
allows to neglect the dissipative effects due to the Coulomb interaction
with the metal (cf. Figs. \ref{fig:im_real_ratio} and \ref{fig:dynamic_conductivity_2D}).
The reason for this lies in the existence of the additional degree
of freedom in the electron gas (perpendicular momentum $k$), which
produces (upon integration) a weakening of dissipation effects. At
this point it is tempting to speculate that a semi-infinite 3D metal,
or a superconducting wire embedded in a 3D normal matrix, would provide
an additional degree of freedom (momentum $k^{\prime}$ perpendicular
to the wire and to $k$), and would weaken further the impact of dissipation
in the metal.

These remarks are relevant to works suggesting the possibility to
 stabilize the superconductivity in 1D systems by coupling them to
a bath of normal quasiparticles\citep{buchler04_sit_finite_length_wire,fu06_stabilization_of_superconductivity_in_nanowires_by_dissipation,lobos09_dissipation_scwires}.
In these works, the basic underlying physical idea is that the normal
bath provides a source of friction for the phase field $\theta\left(\mathbf{x}\right)$
which tends to quench its fluctuations and therefore, to favor superconductivity
(very much like in the case of a resistively shunted Josephson junction\citep{chakravarty82_macroscopic_quantum_tunneling,bray82_macroscopic_quantum_tunneling,schmid83_diffusion_dissipative_quantum_system}).
However, little attention has been given up to now to the simultaneous
dissipative effects induced by the Coulomb interaction with the electrons
in the metal, which produce friction in the dual field $\rho_{s}\left(\mathbf{x}\right)$,
and therefore tends to increase phase fluctuations, thus deteriorating
superconducting properties. In that sense, our results show that the
best condition would be to screen the Coulomb interaction with a clean
(i.e., large diffusion constant $D$) metallic film (rather than a
wire). This result lends credence to the analysis made in Ref. \onlinecite{lobos09_dissipation_scwires},
where %
{}it was assumed that the Coulomb interactions only renormalize the
bare Luttinger parameters of a superconducting wire in contact with
a 2D normal diffusive metal system. %
{}%
{}

Many other issues remain to be addressed to get an accurate physical
description of a superconducting wire coupled to a dissipative electron
gas, such as the aforementioned effect of topological excitations\citep{zaikin97},
localization effects in the gates as a consequence of disorder, simultaneous
effect of Coulomb interactions and Andreev tunneling, etc. We expect
that our results inspire other works along these lines.

%
{}

%
{}

%
{}

%
{}

%
{}

{}

{}%
{}

\begin{acknowledgments}
%
{}This work was supported by the Swiss National Foundation under MaNEP
and division II.
\end{acknowledgments}
\appendix

\section{\label{sec:Derivation_S_eff}Derivation of the effective action}

Although the derivation of the low-energy action for a superconductor
has been studied by several authors\citep{ambegaokar82_josephson_dissipation,zaikin97,vanotterlo98,DePalo99_effective_action_BCS_BEC_crossover},
here we follow more closely the derivation of De Palo \textit{et al.}
\citep{DePalo99_effective_action_BCS_BEC_crossover}. Our starting
point is the decoupling of the interaction terms appearing in $H_{\text{s}}^{0}$
and $H_{\text{int}}$ {[}Eqs. (\ref{eq:h_s}) and (\ref{eq:h_int_nu})
respectively] by the means of Hubbard-Stratonovich transformations
(HSTs)%
{} \begin{widetext} \begin{align}
e^{-\int d\tau\; H_{\text{int}}\left(\tau\right)} & \propto\int\prod_{\nu}\mathcal{D}\left[\tilde{\rho}_{\nu}\right]\; e^{-\frac{1}{2}\sum_{\nu=\pm}\int d^{4}x_{\mu}d^{4}x_{\mu}^{\prime}\;\left[v_{\nu}\left(x_{\mu}-x_{\mu}^{\prime}\right)\right]^{-1}\tilde{\rho}_{\nu}\left(x_{\mu}\right)\tilde{\rho}_{\nu}\left(x_{\mu}^{\prime}\right)+i\int d^{4}x_{\mu}\;\hat{\rho}_{\nu}\left(x_{\mu}\right)\tilde{\rho}_{\nu}\left(x_{\mu}\right)},\label{eq:hs_ph}\end{align}
\begin{align}
e^{\left|U\right|\int d\mathbf{r}d\tau\;\psi_{\uparrow}^{*}\psi_{\downarrow}^{*}\psi_{\downarrow}\psi_{\uparrow}} & \propto\int\mathcal{D}\left[\Delta^{*},\Delta\right]\; e^{-\int d^{4}x_{\mu}\;\frac{\left|\Delta\left(x_{\mu}\right)\right|^{2}}{\left|U\right|}+\int d^{4}x_{\mu}\;\Delta^{*}\left(x_{\mu}\right)\psi_{\downarrow}\left(x_{\mu}\right)\psi_{\uparrow}\left(x_{\mu}\right)+\psi_{\uparrow}^{*}\left(x_{\mu}\right)\psi_{\downarrow}^{*}\left(x_{\mu}\right)\Delta\left(x_{\mu}\right)},\label{eq:hs_pp}\end{align}
\end{widetext}where we have introduced the compact notation $x_{\mu}=\left(\mathbf{r},\tau\right)$
and the bosonic fields $\tilde{\rho}_{\nu}\left(x_{\mu}\right),\Delta^{*}\left(x_{\mu}\right),\Delta\left(x_{\mu}\right)$.
The quantity $\left[v_{\nu}\left(x_{\mu}-x_{\mu}^{\prime}\right)\right]^{-1}$
is a compact notation for the Fourier transform\begin{align}
\left[v_{\nu}\left(x_{\mu}-x_{\mu}^{\prime}\right)\right]^{-1} & =\frac{1}{\beta\Omega}\sum_{k^{\mu}}\frac{e^{ik^{\mu}\left(x_{\mu}-x_{\mu}^{\prime}\right)}}{v_{\nu}\left(k^{\mu}\right)},\label{eq:notation_v**-1}\end{align}
where $v_{\nu}\left(k^{\mu}\right)=\int d^{4}x_{\mu}e^{-ik^{\mu}\left(x_{\mu}-x_{\mu}^{\prime}\right)}v_{\nu}\left(x_{\mu}-x_{\mu}^{\prime}\right)$,
with $v_{\nu}\left(x_{\mu}-x_{\mu}^{\prime}\right)$ the potential
$v_{\nu}\left(x_{\mu}-x_{\mu}^{\prime}\right)\equiv v_{\nu}\left(\mathbf{r}-\mathbf{r}^{\prime}\right)\delta\left(\tau-\tau^{\prime}\right)$.
Note that the mode $k^{\mu}=0$, for which the above HST is formally
ill-defined, can be safely ignored by considering the interaction
with the positive ionic background in the system (not explicitly written
here).

Our next step is to decouple the quadratic term $\tilde{\rho}_{\nu}\left(x_{\mu}\right)\tilde{\rho}_{\nu}\left(x_{\mu}^{\prime}\right)$
in Eq. (\ref{eq:hs_ph}) by the means of an extra HST. According to
Ref. \onlinecite{DePalo99_effective_action_BCS_BEC_crossover}, this has the
advantage of introducing the \textit{physical densities} (symmetric
and antisymmetric) of the problem (cf. Eq. \ref{eq:rho_nu}). Then,
\begin{widetext}\begin{align}
e^{-\frac{1}{2}\int d^{4}x_{\mu}d^{4}x_{\mu}^{\prime}\;\left[v_{\nu}\left(x_{\mu}-x_{\mu}^{\prime}\right)\right]^{-1}\tilde{\rho}_{\nu}\left(x_{\mu}\right)\tilde{\rho}_{\nu}\left(x_{\mu}^{\prime}\right)} & \propto\int\mathcal{D}\left[\rho_{\nu}\right]\; e^{-\frac{1}{2}\int d^{4}x_{\mu}d^{4}x_{\mu}^{\prime}\;\rho_{\nu}\left(x_{\mu}\right)v_{\nu}\left(x_{\mu}-x_{\mu}^{\prime}\right)\rho_{\nu}\left(x_{\mu}^{\prime}\right)-i\int d^{4}x_{\mu}\;\tilde{\rho}_{\nu}\left(x_{\mu}\right)\rho_{\nu}\left(x_{\mu}\right)}.\label{eq:hs_rho}\end{align}
\end{widetext}Note that the formal integration of the field $\tilde{\rho}_{\nu}$
gives the functional-delta function\citep{negele} $\delta\left[\hat{\rho}_{\nu}-\rho_{\nu}\right]$.
This fact allows to interpret the HS fields $\rho_{\nu}$ as the physical
electronic densities\citep{DePalo99_effective_action_BCS_BEC_crossover}.

It is convenient to write the action of the system after these manipulations

\begin{alignat}{1}
S & =\int d^{4}x_{\mu}\;\sum_{\sigma}\left\{ \psi_{\sigma}^{*}\left(\partial_{\tau}-\mu\right)\psi_{\sigma}+\frac{1}{2m}\left[\nabla\psi_{\sigma}^{*}\right]\left[\nabla\psi_{\sigma}\right]\right\} +\nonumber \\
 & +\int d^{4}x_{\mu}\;\left\{ \frac{\left|\Delta\left(x_{\mu}\right)\right|^{2}}{\left|U\right|}-\Delta^{*}\left(x_{\mu}\right)\psi_{\downarrow}\psi_{\uparrow}-\psi_{\uparrow}^{*}\psi_{\downarrow}^{*}\Delta\left(x_{\mu}\right)\right\} +\nonumber \\
 & +\int dx_{\mu}\;\sum_{\sigma}\left\{ \eta_{\sigma}^{*}\left(\partial_{\tau}-\mu+V_{\text{i}}\right)\eta_{\sigma}+\frac{1}{2m}\left[\nabla\eta_{\sigma}^{*}\right]\left[\nabla\eta_{\sigma}\right]\right\} +\nonumber \\
 & +\frac{1}{2}\sum_{\nu=\pm}\int d^{4}x_{\mu}d^{4}x_{\mu}^{\prime}\;\rho_{\nu}\left(x_{\mu}\right)v_{\nu}\left(x_{\mu}-x_{\mu}^{\prime}\right)\rho_{\nu}\left(x_{\mu}^{\prime}\right)+\nonumber \\
 & +i\sum_{\nu=\pm}\int d^{4}x_{\mu}\;\tilde{\rho}_{\nu}\left(x_{\mu}\right)\left[\rho_{\nu}\left(x_{\mu}\right)-\hat{\rho}_{\nu}\left(x_{\mu}\right)\right],\label{eq:S_decoupled}\end{alignat}
where for simplicity we have dropped the arguments in the fermionic
fields $\psi_{\sigma}$ and $\eta_{\sigma}$ and in the disorder potential
$V_{\text{i}}=V_{\text{i}}\left(\mathbf{r}\right)$.

The next step is to perform the saddle-point approximation with respect
to the bosonic fields $\Delta\left(x_{\mu}\right),\Delta^{*}\left(x_{\mu}\right),\tilde{\rho}_{\nu}\left(x_{\mu}\right),\rho_{\nu}\left(x_{\mu}\right)$,
which gives the equations \begin{align}
\frac{\delta S}{\delta\Delta^{*}\left(x_{\mu}\right)} & =0=\frac{\Delta\left(x_{\mu}\right)}{\left|U\right|}-\psi_{\downarrow}\left(x_{\mu}\right)\psi_{\uparrow}\left(x_{\mu}\right),\label{eq:saddle_point_Delta_1}\\
\frac{\delta S}{\delta\Delta\left(x_{\mu}\right)} & =0=\frac{\Delta^{*}\left(x_{\mu}\right)}{\left|U\right|}-\psi_{\uparrow}^{*}\left(x_{\mu}\right)\psi_{\downarrow}^{*}\left(x_{\mu}\right),\label{eq:saddle_point_Delta_2}\\
\frac{\delta S}{\delta\tilde{\rho}_{\nu}\left(x_{\mu}\right)} & =0=\rho_{\nu}\left(x_{\mu}\right)-\hat{\rho}_{\nu}\left(x_{\mu}\right),\label{eq:saddle_point_rho_HS}\\
\frac{\delta S}{\delta\rho_{\nu}\left(x_{\mu}\right)} & =0=\int dx_{\mu}^{\prime}\; v_{\nu}\left(x_{\mu}-x_{\mu}^{\prime}\right)\rho_{\nu}\left(x_{\mu}^{\prime}\right)+i\tilde{\rho}_{\nu}\left(x_{\mu}\right).\label{eq:saddle_point_rho}\end{align}
The first two equations reproduce the well-known BCS gap-equation\citep{Tinkham},
while the other two give the relationship between $\rho_{\nu},\tilde{\rho}_{\nu}$
and the electronic density. These equations provide the starting point
for a controlled expansion in terms of Gaussian fluctuations of the
bosonic fields around the uniform solutions $\Delta^{\left(0\right)}$,
$\rho_{\nu}^{\left(0\right)}$ and $\tilde{\rho}_{\nu}^{\left(0\right)}$.
In what follows, we assume that the values of $\Delta^{\left(0\right)}$,
$\rho_{\nu}^{\left(0\right)}$ and $\tilde{\rho}_{\nu}^{\left(0\right)}$
are known. When these solutions are inserted back into the action
Eq. (\ref{eq:S_decoupled}), we notice that the quantity $\tilde{\rho}_{\nu}^{\left(0\right)}=i\rho_{\nu}^{\left(0\right)}\int d^{4}x_{\mu}\; v_{\nu}\left(x_{\mu}\right)$
can be absorbed in a renormalization of the chemical potential $\mu_{\nu}$
due to the effect of Coulomb interactions, while the divergent quantity
$\frac{1}{2}\sum_{\nu=\pm}\left(\rho_{\nu}^{\left(0\right)}\right)^{2}\int d^{4}x_{\mu}d^{4}x_{\mu}^{\prime}\; v_{\nu}\left(x_{\mu}-x_{\mu}^{\prime}\right)$
exactly cancels the contribution coming from the positive ionic background
(which we have not written explicitly here), by imposing the overall
electroneutrality of the system, and consequently we will drop it
in the following. We also drop the constant term $\beta\Omega\frac{\Delta_{0}^{2}}{\left|U\right|}$,
where $\Omega$ is the volume of the superconducting system.

At sufficiently low energies, amplitude fluctuations of the order
parameter can be neglected, and we can write $\Delta\left(x_{\mu}\right)=\Delta_{0}e^{i\theta\left(x_{\mu}\right)}$,
with a real constant $\Delta_{0}=\left|\Delta^{\left(0\right)}\right|$.
We can absorbe the phase field by the means of a transformation of
the fermion field\begin{align*}
\psi_{\sigma}\left(x_{\mu}\right) & \rightarrow\psi_{\sigma}^{\prime}\left(x_{\mu}\right)=\psi_{\sigma}\left(x_{\mu}\right)e^{i\theta\left(x_{\mu}\right)/2}.\end{align*}

The expression of the effective action is considerably simplified
introducing the Nambu notation\begin{align*}
\boldsymbol{\Psi}\left(x_{\mu}\right) & \equiv\left(\begin{array}{c}
\psi_{\uparrow}\left(x_{\mu}\right)\\
\psi_{\downarrow}^{*}\left(x_{\mu}\right)\end{array}\right), & \boldsymbol{\eta}\left(x_{\mu}\right) & \equiv\left(\begin{array}{c}
\eta_{\uparrow}\left(x_{\mu}\right)\\
\eta_{\downarrow}^{*}\left(x_{\mu}\right)\end{array}\right),\end{align*}
which allows to write the action as%
{} \begin{align*}
S & \simeq\int d^{4}x_{\mu}\;\left\{ \boldsymbol{\Psi}^{\dagger}\left[\boldsymbol{\mathcal{A}}_{0,s}-\boldsymbol{\Sigma}_{s}\right]\boldsymbol{\Psi}+\boldsymbol{\eta}^{\dagger}\left[\boldsymbol{\mathcal{A}}_{0,n}-\boldsymbol{\Sigma}_{n}\right]\boldsymbol{\eta}\right\} +\\
 & +\frac{1}{2}\sum_{\nu}\int d^{4}x_{\mu}d^{4}x_{\mu}^{\prime}\;\delta\rho_{\nu}\left(x_{\mu}\right)v_{\nu}\left(x_{\mu}-x_{\mu}^{\prime}\right)\delta\rho_{\nu}\left(x_{\mu}^{\prime}\right)+\\
 & +i\sum_{\nu=\pm}\int d^{4}x_{\mu}\;\delta\tilde{\rho}_{\nu}\left(x_{\mu}\right)\left[\rho_{\nu}^{\left(0\right)}+\delta\rho_{\nu}\left(x_{\mu}\right)\right].\end{align*}
where\begin{align}
\delta\tilde{\rho}_{\nu}\left(x_{\mu}\right) & \equiv\tilde{\rho}_{\nu}\left(x_{\mu}\right)-\tilde{\rho}_{\nu}^{\left(0\right)},\label{eq:rho_tilde_nu_fluctuation}\\
\delta\rho_{\nu}\left(x_{\mu}\right) & \equiv\rho_{\nu}\left(x_{\mu}\right)-\rho_{\nu}^{\left(0\right)},\label{eq:rho_nu_fluctuation}\end{align}
are the fluctuations of the density around the saddle-point solutions,
and \begin{align*}
\boldsymbol{\mathcal{A}}_{0,s} & \equiv\left\{ \partial_{\tau}\right\} \boldsymbol{\hat{\tau}}_{0}-\left\{ \frac{\nabla^{2}}{2m}+\mu_{s}\right\} \boldsymbol{\hat{\tau}}_{3}-\Delta_{0}\boldsymbol{\hat{\tau}}_{1},\\
\boldsymbol{\Sigma}_{s} & \equiv-\left\{ \frac{i\left(\partial_{\tau}\theta\right)}{2}+\frac{\left(\nabla\theta\right)^{2}}{8m}-i\sum_{\nu}\delta\tilde{\rho}_{\nu}\right\} \boldsymbol{\hat{\tau}}_{3}+\frac{i\left(\nabla\theta\right)\nabla}{2m}\boldsymbol{\hat{\tau}}_{0},\\
\boldsymbol{\mathcal{A}}_{0,n} & \equiv\left\{ \partial_{\tau}\right\} \boldsymbol{\hat{\tau}}_{0}-\left\{ \frac{\nabla^{2}}{2m}+\mu_{n}\right\} \boldsymbol{\hat{\tau}}_{3},\\
\boldsymbol{\Sigma}_{n} & \equiv i\sum_{\nu}\left(\nu\right)\delta\tilde{\rho}_{\nu}\boldsymbol{\hat{\tau}}_{3},\end{align*}
where $\boldsymbol{\hat{\tau}}_{i}$ are the Pauli matrices and where
we have used the fact that $\frac{i\nabla\theta\left[\overrightarrow{\nabla}-\overleftarrow{\nabla}\right]}{2m}=\frac{i\nabla\theta\nabla}{m}$
in a translationally invariant system.

{}

The next step consists in using the expansion formula \begin{align*}
\text{Tr ln}\left[\boldsymbol{\mathcal{A}}_{0}-\boldsymbol{\Sigma}\right] & =\text{Tr ln }\boldsymbol{\mathcal{A}}_{0}-\sum_{n=1}^{\infty}\frac{\left(-1\right)^{n}}{n}\text{Tr }\left[\mathbf{G}_{0}\boldsymbol{\Sigma}\right]^{n},\end{align*}
where $\mathbf{G}_{0}\equiv-\left[\boldsymbol{\mathcal{A}}_{0}\right]^{-1}$.
Truncating the series at second order (i.e., Gaussian fluctuations),
we obtain \begin{align*}
S & \simeq-\text{Tr }\left[\mathbf{G}_{0,s}\boldsymbol{\Sigma}_{s}\right]+\frac{1}{2}\text{Tr }\left[\mathbf{G}_{0,s}\boldsymbol{\Sigma}_{s}\right]^{2}+\\
 & -\text{Tr }\left[\mathbf{G}_{0,n}\boldsymbol{\Sigma}_{n}\right]+\frac{1}{2}\text{Tr }\left[\mathbf{G}_{0,n}\boldsymbol{\Sigma}_{n}\right]^{2}+\\
 & +\frac{1}{2}\sum_{\nu}\int d^{4}x_{\mu}d^{4}x_{\mu}^{\prime}\;\delta\rho_{\nu}\left(x_{\mu}\right)v_{\nu}\left(x_{\mu}-x_{\mu}^{\prime}\right)\delta\rho_{\nu}\left(x_{\mu}^{\prime}\right)+\\
 & +i\sum_{\nu=\pm}\int d^{4}x_{\mu}\;\delta\tilde{\rho}_{\nu}\left(x_{\mu}\right)\left[\rho_{\nu}^{\left(0\right)}+\delta\rho_{\nu}\left(x_{\mu}\right)\right],\end{align*}
where the propagators in Nambu space $\mathbf{G}_{0,s}$ and $\mathbf{G}_{0,n}$
write\begin{align*}
\mathbf{G}_{0,s} & =\left(\begin{array}{cc}
g_{0,s}\left(x_{\mu}\right) & f_{0,s}\left(x_{\mu}\right)\\
\bar{f}_{0,s}\left(x_{\mu}\right) & \bar{g}_{0,s}\left(x_{\mu}\right)\end{array}\right),\\
\mathbf{G}_{0,n} & =\left(\begin{array}{cc}
g_{0,n}\left(x_{\mu}\right) & 0\\
0 & \bar{g}_{0,n}\left(x_{\mu}\right)\end{array}\right),\end{align*}
and where $g_{0,s}\left(x_{\mu}\right)\equiv-\left\langle T_{\tau}\psi_{\uparrow}\left(x_{\mu}\right)\psi_{\uparrow}^{*}\left(0\right)\right\rangle $
and $\bar{g}_{0,s}\left(x_{\mu}\right)\equiv\left\langle T_{\tau}\psi_{\downarrow}^{*}\left(x_{\mu}\right)\psi_{\downarrow}\left(0\right)\right\rangle $
denote respectively the particle and hole propagators in the superconductor,
while $f_{0,s}\left(x_{\mu}\right)\equiv\left\langle T_{\tau}\psi_{\downarrow}\left(x_{\mu}\right)\psi_{\uparrow}\left(0\right)\right\rangle $,
$\bar{f}_{0,s}\left(x_{\mu}\right)\equiv\left\langle T_{\tau}\psi_{\uparrow}^{*}\left(x_{\mu}\right)\psi_{\downarrow}^{*}\left(0\right)\right\rangle $
are the anomalous ones\citep{fetter}. Similarly $g_{0,n}\left(x_{\mu}\right)\equiv-\left\langle T_{\tau}\eta_{\uparrow}\left(x_{\mu}\right)\eta_{\uparrow}^{*}\left(0\right)\right\rangle $
and $\bar{g}_{0,n}\left(x_{\mu}\right)\equiv\left\langle T_{\tau}\eta_{\downarrow}^{*}\left(x_{\mu}\right)\eta_{\downarrow}\left(0\right)\right\rangle $
are the particle and hole propagators in the normal metal, respectively.

{}The evaluation of the traces yields\begin{align*}
\text{Tr}\left[\mathbf{G}_{0,s}\boldsymbol{\Sigma}_{s}\right] & =-\rho_{s}^{\left(0\right)}\int d^{4}x_{\mu}\;\left[\frac{i}{2}\partial_{\tau}\theta+\frac{\left(\nabla\theta\right)^{2}}{8m}-i\sum_{\nu}\delta\tilde{\rho}_{\nu}\right]_{x_{\mu}},\\
\text{Tr}\left[\mathbf{G}_{0,s}\boldsymbol{\Sigma}_{s}\right]^{2} & =\int d^{4}x_{\mu}d^{4}x_{\mu}^{\prime}\;\biggl\{\chi_{0,s}\left(x_{\mu}-x_{\mu}^{\prime}\right)\times\\
 & \times\left[\frac{1}{2}\partial_{\tau}\theta-\sum_{\nu}\delta\tilde{\rho}_{\nu}\right]_{x_{\mu}}\left[\frac{1}{2}\partial_{\tau}\theta-\sum_{\nu^{\prime}}\delta\tilde{\rho}_{\nu^{\prime}}\right]_{x_{\mu}^{\prime}}+\\
 & -\mathcal{D}^{\prime}\left(x_{\mu}-x_{\mu}^{\prime}\right)\left[\nabla\theta\right]_{x_{\mu}}\left[\nabla\theta\right]_{x_{\mu}^{\prime}}\biggr\},\\
\text{Tr}\left[\mathbf{G}_{0,n}\boldsymbol{\Sigma}_{n}\right] & =i\rho_{n}^{\left(0\right)}\int d^{4}x_{\mu}\;\left[\sum_{\nu}\left(\nu\right)\delta\tilde{\rho}_{\nu}\right]_{x_{\mu}},\\
\text{Tr}\left[\mathbf{G}_{0,n}\boldsymbol{\Sigma}_{n}\right]^{2} & =\int d^{4}x_{\mu}d^{4}x_{\mu}^{\prime}\;\chi_{0,n}\left(x_{\mu}-x_{\mu}^{\prime}\right)\times\\
 & \times\Biggl[\sum_{\nu}\left(\nu\right)\delta\tilde{\rho}_{\nu}\Biggr]_{x_{\mu}}\Biggl[\sum_{\nu^{\prime}}\left(\nu^{\prime}\right)\delta\tilde{\rho}_{\nu^{\prime}}\Biggr]_{x_{\mu}^{\prime}},\end{align*}
where for simplicity we have used the campact notation $\left[\mathcal{A}\right]_{x_{\mu}}\equiv\mathcal{A}\left(x_{\mu}\right)$,
and where we have defined \begin{align*}
\mathcal{D}^{\prime}\left(x_{\mu}-x_{\mu}^{\prime}\right) & \equiv\frac{1}{2m^{2}}\left[\nabla g_{0,s}\left(x_{\mu}^{\prime}-x_{\mu}\right)\nabla g_{0,s}\left(x_{\mu}-x_{\mu}^{\prime}\right)+\right.\\
 & \left.+\nabla f_{0,s}\left(x_{\mu}^{\prime}-x_{\mu}\right)\nabla f_{0,s}\left(x_{\mu}-x_{\mu}^{\prime}\right)\right],\end{align*}
and the density-density correlation functions\begin{align}
\chi_{0,s}\left(x_{\mu}-x_{\mu}^{\prime}\right) & \equiv-\left\langle T_{\tau}\delta\rho_{s}\left(x_{\mu}\right)\delta\rho_{s}\left(x_{\mu}^{\prime}\right)\right\rangle ,\nonumber \\
 & =-2g_{0,s}\left(x_{\mu}-x_{\mu}^{\prime}\right)g_{0,s}\left(x_{\mu}^{\prime}-x_{\mu}\right)+\nonumber \\
 & +2f_{0,s}\left(x_{\mu}-x_{\mu}^{\prime}\right)f_{0,s}\left(x_{\mu}^{\prime}-x_{\mu}\right),\label{eq:sc_susceptibility}\\
\chi_{0,n}\left(x_{\mu}-x_{\mu}^{\prime}\right) & \equiv-\left\langle T_{\tau}\delta\rho_{n}\left(x_{\mu}\right)\delta\rho_{n}\left(x_{\mu}^{\prime}\right)\right\rangle ,\nonumber \\
 & =-2g_{0,n}\left(x_{\mu}-x_{\mu}^{\prime}\right)g_{0,n}\left(x_{\mu}^{\prime}-x_{\mu}\right).\label{eq:n_susceptibility}\end{align}
{}

The final step is to integrate out the modes $\delta\tilde{\rho}_{\nu}\left(x_{\mu}\right)$.
To that aim, we decouple the mixed term $\sim\left(\delta\tilde{\rho}_{\nu}\right)\left(\delta\tilde{\rho}_{\bar{\nu}}\right)^{\prime}$
appearing in $\text{Tr}\left[\mathbf{G}_{0,s}\boldsymbol{\Sigma}_{s}\right]^{2}$
and $\text{Tr}\left[\mathbf{G}_{0,n}\boldsymbol{\Sigma}_{n}\right]^{2}$
by returning to the original representation for the densities {[}cf.
Eq. (\ref{eq:rho_nu})]\begin{align*}
\delta\tilde{\rho}_{s} & =\frac{\delta\tilde{\rho}_{+}+\delta\tilde{\rho}_{-}}{2},\\
\delta\tilde{\rho}_{n} & =\frac{\delta\tilde{\rho}_{+}-\delta\tilde{\rho}_{-}}{2},\end{align*}
and integrate out the fields $\delta\tilde{\rho}_{n}\left(x_{\mu}\right)$
and $\delta\tilde{\rho}_{s}\left(x_{\mu}\right)$ instead. %
{} From here we see that the term $\sim\left(\partial_{\tau}\theta\right)\left(\partial_{\tau}\theta\right)^{\prime}$
cancels, as in Ref. \onlinecite{DePalo99_effective_action_BCS_BEC_crossover}.
We finally obtain

\begin{align}
S & =\frac{i}{2}\int d^{4}x_{\mu}\;\partial_{\tau}\theta\left(x_{\mu}\right)\rho_{s}\left(x_{\mu}\right)+\nonumber \\
 & +\int d^{4}x_{\mu}d^{4}x_{\mu}^{\prime}\;\left\{ \frac{1}{2}\mathcal{D}\left(x_{\mu}-x_{\mu}^{\prime}\right)\nabla\theta\left(x_{\mu}\right)\nabla\theta\left(x_{\mu}^{\prime}\right)+\right.\nonumber \\
 & \left.+\boldsymbol{\delta\rho}^{\dagger}\left(x_{\mu}\right)\boldsymbol{V}\left(x_{\mu}-x_{\mu}^{\prime}\right)\boldsymbol{\delta\rho}\left(x_{\mu}^{\prime}\right)\right\} \label{eq:S_eff_dddd}\end{align}
where we have defined\begin{align}
\mathcal{D}\left(x_{\mu}\right) & \equiv\frac{\rho_{s}^{\left(0\right)}}{4m}\delta\left(x_{\mu}\right)-\mathcal{D}^{\prime}\left(x_{\mu}\right),\label{eq:sc_stiffness}\\
\boldsymbol{\delta\rho}\left(x_{\mu}\right) & \equiv\left(\begin{array}{c}
\delta\rho_{s}\left(x_{\mu}\right)\\
\delta\rho_{n}\left(x_{\mu}\right)\end{array}\right),\nonumber \\
\boldsymbol{V}\left(x_{\mu}\right) & \equiv\left(\begin{array}{cc}
\left[\chi_{0,s}\left(x_{\mu}\right)\right]^{-1} & 0\\
0 & \left[\chi_{0,n}\left(x_{\mu}\right)\right]^{-1}\end{array}\right)+\nonumber \\
 & +\delta\left(\tau\right)\left(\begin{array}{cc}
v\left(x_{\mu},0\right) & v\left(x_{\mu},d\right)\\
v\left(x_{\mu},d\right) & v\left(x_{\mu},0\right)\end{array}\right),\nonumber \end{align}
where again we used a compact notation of Eq. (\ref{eq:notation_v**-1}).

\section{\label{sec:Fourier_Transforms}Density susceptibility and superconducting
stiffness in the limit $q\rightarrow0$}

{}From Eq. (\ref{eq:sc_susceptibility}), the Fourier transforms reads
\begin{align*}
\chi_{0,s}\left(q^{\mu}\right) & =\frac{2}{\beta\Omega}\sum_{k^{\mu}}\left[-g_{0,s}\left(k^{\mu}\right)g_{0,s}\left(k^{\mu}-q^{\mu}\right)+\right.\\
 & \left.+f_{0,s}\left(k^{\mu}\right)f_{0,s}\left(k-q^{\mu}\right)\right].\end{align*}
In the limit $q^{\mu}\rightarrow0$, we obtain\begin{align*}
\lim_{q^{\mu}\rightarrow0}\chi_{0,s}\left(q^{\mu}\right) & \rightarrow-\frac{2}{\Omega}\sum_{\mathbf{k}}\frac{1}{\beta}\sum_{n}\frac{\left(i\nu_{n}\right)^{2}+\xi_{\mathbf{k}}^{2}-\Delta_{0}^{2}}{\left[\left(i\nu_{n}\right)^{2}-E_{\mathbf{k}}^{2}\right]^{2}}\\
 & =-\frac{2}{\Omega}\left[\sum_{k}\frac{n_{F}\left(E_{\mathbf{k}}\right)}{2E_{\mathbf{k}}}-\frac{n_{F}\left(-E_{\mathbf{k}}\right)}{2E_{\mathbf{k}}}\right],\end{align*}
with $\xi_{\mathbf{k}}\equiv\frac{k^{2}}{2m}-\mu_{s}$, and $E_{\mathbf{k}}\equiv\sqrt{\xi_{\mathbf{k}}^{2}+\Delta_{0}^{2}}$.
At $T=0$\begin{align}
\lim_{q^{\mu}\rightarrow0}\chi_{0,s}\left(q^{\mu}\right) & =\frac{2}{\Omega}\sum_{\mathbf{k}}\frac{1}{2E_{\mathbf{k}}}.\nonumber \\
 & =\mathcal{N}_{s}^{\left(0\right)}\gamma,\label{eq:susceptibility_sc_limit_q_0}\end{align}
where $\gamma\equiv\int_{-\omega_{D}}^{\omega_{D}}d\xi\frac{1}{\sqrt{\xi^{2}+\Delta_{0}^{2}}}=\ln\left[\frac{\omega_{D}+\sqrt{\omega_{D}^{2}+\Delta_{0}^{2}}}{-\omega_{D}+\sqrt{\omega_{D}^{2}+\Delta_{0}^{2}}}\right]\approx2\ln\left[\frac{2\omega_{D}}{\Delta_{0}}\right]$,
and where $\omega_{D}$ is a high-energy cutoff.

{}Similarly, the Fourier transform of the superconducting stiffness
reads \begin{align*}
\mathcal{D}\left(q^{\mu}\right) & \equiv\frac{\rho_{s}^{\left(0\right)}}{4m}-\mathcal{D}^{\prime}\left(q^{\mu}\right),\end{align*}
with \begin{align*}
\mathcal{D}^{\prime}\left(q^{\mu}\right) & \equiv\frac{1}{\Omega}\sum_{k^{\mu}}\frac{-\mathbf{k}.\left(\mathbf{k}-\mathbf{q}\right)}{2m^{2}}\times\\
 & \times\left[f\left(k^{\mu}\right)f\left(k^{\mu}-q^{\mu}\right)+g\left(k^{\mu}\right)g\left(k^{\mu}-q^{\mu}\right)\right]\\
 & =\frac{1}{V}\sum_{\mathbf{k}}\frac{-\mathbf{k}.\left(\mathbf{k}-\mathbf{q}\right)}{2m^{2}}\times\\
 & \times\frac{1}{\beta}\sum_{n}\frac{\Delta_{0}^{2}+\left(i\nu_{n}+\xi_{\mathbf{k}}\right)\left(i\nu_{n}-i\omega_{m}+\xi_{\mathbf{k}-\mathbf{q}}\right)}{\left[\left(i\nu_{n}\right)^{2}-E_{\mathbf{k}}^{2}\right]\left[\left(i\nu_{n}-i\omega_{m}\right)^{2}-E_{\mathbf{k}-\mathbf{q}}^{2}\right]},\end{align*}
Evaluating the Matsubara sum over the fermionic frequencies $i\nu_{n}$,
we obtain the result in the limit $q^{\mu}\rightarrow0$\begin{align*}
\lim_{q^{\mu}\rightarrow0}\mathcal{D}^{\prime}\left(q^{\mu}\right) & \approx-\frac{1}{V}\sum_{\mathbf{k}}\frac{k^{2}}{2m^{2}}\frac{n_{F}\left(E_{\mathbf{k}}\right)-n_{F}\left(E_{\mathbf{k}-\mathbf{q}}\right)}{E_{\mathbf{k}}-E_{\mathbf{k}-\mathbf{q}}}\\
 & \approx-\frac{1}{V}\sum_{\mathbf{k}}\frac{k^{2}}{2m^{2}}\frac{\partial n_{F}\left(E_{\mathbf{k}}\right)}{\partial E_{\mathbf{k}}},\end{align*}
which vanishes in the limit $T\rightarrow0$, and we recover the well-know
result\citep{Tinkham,fetter} \begin{align*}
\lim_{q^{\mu}\rightarrow0}\mathcal{D}\left(q^{\mu}\right) & \equiv\mathcal{D}_{0}=\frac{\rho_{s}^{\left(0\right)}}{4m}.\end{align*}

\bibliographystyle{apsrev}

\begin{thebibliography}{62}
\bibitem[{not()}]{note_technical_difficuly_dynamic_conductivity}
\bibinfo{note}{Note, however, that in general is technically difficult to have
  complete experimental access to $\sigma\left(q,\omega\right)$ as a function
  of $q$ and $\omega$ since, typically, optical measurements allow only to
  measure $\sigma\left(q=0,\omega\right)$.}
\end{thebibliography}

\end{document}